\documentclass[aip,amsmath,amssymb,reprint]{revtex4-1}
\usepackage{graphicx}
\usepackage{dcolumn}
\usepackage{bm}
\usepackage{bookmark}
\usepackage{xcolor}
\usepackage[utf8]{inputenc}
\usepackage[T1]{fontenc}
\usepackage{mathptmx}
\usepackage{etoolbox}
\newcommand\uvec[1]{{\ensuremath{\hat{\bm{#1}}}}}

\graphicspath{{./},{./figs/}}

\makeatletter
\def\@email#1#2{%
 \endgroup
 \patchcmd{\titleblock@produce}
  {\frontmatter@RRAPformat}
  {\frontmatter@RRAPformat{\produce@RRAP{*#1\href{mailto:#2}{#2}}}\frontmatter@RRAPformat}
  {}{}
}%
\makeatother

\begin{document}

\title{Dynamics of microswimmers near a liquid-liquid interface with viscosity difference}

\author{Chao Feng}
    \affiliation{Department of Chemical Engineering, Kyoto University, Kyoto 615-8510, Japan}
\author{John J. Molina}
    \affiliation{Department of Chemical Engineering, Kyoto University, Kyoto 615-8510, Japan}
\author{Matthew S. Turner}
    \affiliation{Department of Chemical Engineering, Kyoto University, Kyoto 615-8510, Japan}
    \affiliation{Department of Physics, University of Warwick, Coventry CV4 7AL, UK}
\author{Ryoichi Yamamoto}
    \affiliation{Department of Chemical Engineering, Kyoto University, Kyoto 615-8510, Japan}
    \email{ryoichi@cheme.kyoto-u.ac.jp}

\date{\today}

\begin{abstract}
Transport of material across liquid interfaces is ubiquitous for living cells and is also a crucial step in drug delivery and in many industrial processes. The fluids that are present on either side of the interfaces will usually have different viscosities. We present a physical model for the dynamics of microswimmers near a soft and penetrable interface that we solve using computer simulations of Navier-Stokes flows. The literature contains studies of similar isoviscous fluid systems, where the two fluids have the same viscosity. Here we extend this to the more general case where they have different viscosities. We investigate the effect of the fluid viscosity ratio on the movement patterns of microswimmers. We find that swimmers systematically reorientate towards the region containing the lower viscosity fluid. Ultimately this is expected to drive the swimmers to behave as if they are more inclined to swim in low viscosity fluids. 

Furthermore, in addition to the types of swimming already reported in the isoviscous system, i.e. bouncing, sliding and penetrating, we observed a hovering motion, in which strong pullers swim parallel to the interface with a certain distance, which is consistent with the dynamics of such swimmers near the solid wall.
\end{abstract}

\maketitle

\section{introduction}

Across the natural world, it is common for micro swimmers to navigate through complex fluid environments, e.g., spermatozoa in seminal plasma and cervical mucus. \cite{kirkman2011sperm,gadelha2010nonlinear,bjorndahl2010usefulness}. Trans-membrane transport of bacteria and viruses is also a key stage in infection. \cite{nguyen2020targeted}
Therefore, research into the dynamics of swimmers in such environments is of vital importance to understand many biological transport processes, as well as for the development of potential future artificial micro-machines, which could be used for targeted delivery in complex fluid environments.\cite{ceylan20183d,mostaghaci2017bioadhesive,park2017multifunctional,yan2015magnetite,laage2006molecular}

Most studies on the dynamics of microswimmers in inhomogeneous multi-phase systems has focused on swimming in the vicinity of boundaries, mainly solid-fluid boundaries\cite{ishimoto2013squirmer,lintuvuori2016hydrodynamic,li2014hydrodynamic,pagonabarraga2013structure,shen2018hydrodynamic,volpe2011microswimmers,fadda2020dynamics,lauga2006swimming} and fluid-air interfaces\cite{ishimoto2013squirmer,di2011swimming}.
For example, Lauga et al. \cite{lauga2006swimming} showed that E. coli was shown to exhibit a clockwise circular swimming motion near a solid-fluid boundary, whereas Leonardo et al.\cite{di2011swimming} report a counterclockwise rotation near a free surface, such as the fluid-air interface. In both cases, this circular motion can be explained as arising from the hydrodynamic interactions. 

However, few theoretical studies have attempted to construct a general hydrodynamic description of microswimmers near (complex) fluid-fluid interfaces. This may be due  to the high computational costs associated with treating deformable and penetrable boundaries. Of particular (biological) importance is the case where fluids have mismatched viscocities. This can dramatically affect the swimmer dynamics. For example, spermatozoa exhibit very different tail wave forms, depending on the viscosity of the fluid\cite{kirkman2011sperm,gaffney2011mammalian}. 
Sperm in low viscosity medium have a significant side-to-side movement across the directional axis. 
However, in high viscosity fluids, e.g., mucus, such head yaw is reduced by using a distinct ``meandering'' waveform with less lateral movement across the directional axis. In this way, the spermatozoa is able to swim with approximately the same velocity in either fluid, saline or mucus, despite the large viscosity difference between the two.

Previously, we have studied the dynamics of microswimmers near a soft, deformable and penetrable interface in an isoviscous system\cite{feng2022dynamics}. In this paper, we extend our work to consider the dynamics of swimmers at the interface of two fluids with mismatched viscosities. First, we review the computational model we have used, and detail how it is extended to take the variable viscosity into account. Then, we analyse the effect of the viscosity on the motion of the swimmer in the low Reynolds-number regime. As we previously found for the isoviscous case, different initial trajectories can lead to motion that we characterise as ``bouncing'', ``sliding'', and ``penetrating''. In addition to these modes we also observe a new dynamical ``hovering'' mode, in which the swimmer tends to move parallel to the interface, at a fixed (non-zero) distance. 
By analysing the time evolution of the positions and orientations, we found several sets of trajectories for the bouncing and penetrating motions that exhibit time-reversal symmetry due to the pushers/puller duality.
Compared to the isoviscous case, we find that a viscosity difference can significantly affect the dynamics during collision with the interface, i.e. the relationship between incoming and outgoing angles. Finally, we provide an analysis of 
the swimmer dynamics in the hovering motion. Interestingly, both the trajectories and the time evolution of the orientation are similar to those of swimmers near a solid wall.

\section{simulation methods}
\subsection{The squirmer model}
We model swimmers as spherical squirmers, one of the most widely employed mathematical models for microswimmers\cite{Lighthill,downton2009simulation}.
Thus, swimmers are represented as rigid spherical particles, with a modified
stick boundary condition at their surface. To simplify the description, we only consider the tangential components of the surface velocity, ignoring both azimuthal and radial contributions, which is the usual approximation\cite{pak2014generalized,ishikawa2006hydrodynamic}. The surface slip velocity is given as
\begin{equation}
    {\bm{u}} ^s( \vartheta
    )=\sum_{n=1}^{\infty} \frac{2}{n(n+1)} B_{n}P^{'}_{n}( \cos \vartheta) \sin \vartheta {\bm {\hat{\vartheta}}}, \label{eq:sq}
\end{equation}
where $\vartheta = \cos^{-1} ({\bm{\hat{r}}} \cdot {\bm{\hat{e}}})$ is the polar angle between the swimming direction $\bm {\hat{e}}$ and ${\bm {\hat{r}}} $, a unit vector directed from the center of the squirmer toward the corresponding point on its surface and
${\bm{\hat{\vartheta}}}$ is the unit vector orthogonal to ${\bm{\hat{r}}}$.
$P^{'}_{n}$ is the derivative of the Legendre polynomial of the $n$-th order,
and $B_{n}$ is the magnitude of each mode.

Only the first two modes in Eq.~\ref{eq:sq} are retained:
\begin{equation}
    {\bm{u}} ^s( {\vartheta} )=B_1(\sin \vartheta + \frac{\beta} {2} \sin {2\vartheta} ){\bm {\hat{\vartheta}}}, \label{eq:sq2}
\end{equation}
The coefficient of the first term in Eq.~\ref{eq:sq2}, $B_1$, determines the steady-state swimming velocity of the squirmer $U_0=2/3B_1$.
The coefficient of the second mode, $B_2$, determines the stress exerted by the particles on the fluid.
The ratio $\beta=B_2/B_1$ determines the swimming type and strength.
When $\beta$ is negative, the squirmer is a pusher, which swims generating an extensile flow field (e.g., E. coli);
when $\beta$ is positive, the squirmer is a puller, which swims generating a contractile flow field (e.g., C. reinhardtii). 
The marginal case of $\beta=0$ corresponds to a neutral particle, which is accompanied by a potential flow (e.g., Volvox).
In what follows, we will refer to squirmers with $|\beta|\le1$ as being weak, those with $|\beta|\ge4$ as being strong.

\subsection{Smoothed profile method for binary fluids} \label{SPM}
To simulate the dynamics of particles dispersed in an immiscible binary $A/B$ fluid system, while fully accounting for the hydrodynamic interactions, we consider the coupled equations of motion for the solid particles and the component fluids within the model-H representation, i.e., the Newton-Euler and Cahn-Hilliard Navier-Stokes equations \cite{arai2020direct,lecrivain2020eulerian}. Furthermore, to allow for efficient calculations of many-particle systems, while still providing an accurate description of the many-body hydrodynamic interactions, we employ the Smooth Profile (SP) method\cite{yamamoto2021smoothed}.  Within this approximation, the sharp particle boundaries are replaced by diffuse interfaces of finite thickness. The solid particle domains are thus defined using a continuous order parameter or phase-field $\phi$, defined over the whole computational domain. The $A$/$B$ fluid phases are likewise defined in terms of the corresponding $A$/$B$ order-parameters $\psi_A$ and $\psi_B$. This allows us to easily couple the rigid-body dynamics to the dynamics of the (phase-separating) fluids. In what follows we briefly describe how to solve for the (coupled) rigid-body dynamics, phase-separating dynamics, and fluid dynamics. Detailed descriptions of the SP method and its implementation can be found in our earlier publications.\cite{yamamoto2004smooth,PhysRevE.71.036707,molina2013direct,yamamoto2021smoothed}
\subsubsection{Particle Dynamics}
 The rigid particle dynamics are determined by the Newton--Euler equations of motion (assuming spherical particles):
\begin{align}
        \dot{\bm{R}_i} &=\bm{V}_i,      \label{eq:NE}\\
        \dot{\bm{Q}_i} &={\rm skew}(\bm{\Omega}_i)\cdot\bm{Q}_i,     \\
        {M}_i\dot{\bm{V}_i} &=\bm{F}^H_i+\bm{F}^C_i+\bm{F}^{ext}_i, \\
        \bm{I}_i\cdot\dot{\bm{\Omega}_i}  &=\bm{N}^H_i+\bm{N}^{ext}_i,
\end{align}
where $\bm{R}_i$, $\bm{Q}_i$, $\bm{V}_i$, $\bm{\Omega}_i$ are the positions, orientation matrices, velocities, and angular velocities of particle $i$, respectively; $M_i$ are the masses, and $\bm{I}_i=2/5{M_i}{a_i}^2\bm{\mathsf{I}}$ the moments of inertia for spheres of radius $a_i$ ($\bm{\mathsf{I}}$ the unit tensor);
$\mathrm{skew}(\bm{\Omega}_i)$ is the skew-symmetric matrix for the angular velocity.
The hydrodynamic forces and torques are given by $\bm{F}^H_i$ and $\bm{N}^H_i$, $\bm{F}^C_i$ represents direct particle--particle interactions ($\bm{N}^C_i = 0$), and
$\bm{F}^{ext}_i$ and $\bm{N}^{ext}_i$ are the external forces and torques, respectively.

Within the SP method, the particle domain is accounted for by an order parameter $\phi(\bm{r})$, which can be interpreted as the volume fraction of the solid component in the system
\begin{align}
\phi(\bm{r}) &= \sum_i \phi_i(\bm{r}),
\end{align}
where $\phi_i$ is the phase-field for particle $i$.
This particle phase-field is defined such that it is equal to $1$ in the solid domain, $0$ in the fluid domain, and smoothly interpolates between the two domains across the interfaces (of width $\xi_p$). In this way, the boundaries can be represented through the gradient of the phase-field.
We can then define the velocity field for the particle domain as
\begin{align}
            \phi\bm{u}_p=\sum_i{\phi_i}[\bm{V}_i+\bm{\Omega}_i\times\bm{R}_i].
\end{align}

\subsubsection{Phase-Separating Dynamics}
The order parameters for the A and B phases, $\psi_A ({\bf r})$ and $\psi_B({\bm{r}})$, represent the volume fractions of the constituent components ( $0 \le \phi_\alpha \le 1$). Furthermore, since the sum total of the volume fraction of all components (fluids and particle) is constrained to be unity,
\begin{equation}
        \psi_A+\psi_B+\phi = 1,
\end{equation}
the composition of the A/B phase-separating fluid can be described in terms of a single order parameter $\psi(\bm{r})$, 
\begin{equation}
        \psi = \psi_A-\psi_B.
        \label{eq:psi}
\end{equation}
This order parameter $\psi(\bm{r})$  is defined to be equal to $1$ in the  $A$ domain and $-1$ in the $B$ domain. 

The dynamics for $\psi(\bm{r})$ is determined by the following modified Cahn-Hilliard equation
\begin{align}
        \frac{\partial \psi}{\partial t} + (\bm{u}\cdot \bm{\nabla}) \psi & = \kappa \nabla ^2 \mu_\psi,
        \label{eq:cahn_hilliard}
\end{align}
where $\bm{u}$ is the total velocity field, $\kappa$ is the mobility coefficient, and $\mu_\psi=\delta \mathcal{F}/\delta \psi$ is the chemical potential associated with the order parameter $\psi$. In what follows we will also need to account for the fluid-particle interactions using a second chemical potential associated with the $\phi$ order parameter, $\mu_\phi = \delta \mathcal{F} / \delta \phi$. These chemical potentials are derived from the Ginzburg-Landau free energy functional
\begin{align}
        \mathcal{F}[\psi,\phi] &= \int \mathrm{d}\bm{r} \left[f(\psi) + \frac{\alpha}{2} (\bm{\nabla} \psi)^{2} + w \xi_p \psi (\bm{\nabla}\phi)^{2}\right]
        \label{eq:freeenergy}
\end{align}
The first term in the integrand of Eq.~\ref{eq:freeenergy}, $f( \psi )=\frac{1}{4} \psi^4 - \frac{1}{2} \psi^2$ represents the Landau double-well potential, with two minima at $\psi=1$ and $-1$. The second term is the potential energy associated with the fluid $A$/$B$ interface, the
third term represents the particles' affinity for each of the fluid $A$/$B$ phases. The chemical potentials corresponding to this free-energy functional are then
\begin{align}
   \mu_\psi&=f'(\psi)+\alpha{\nabla}^2\psi + w \xi_p (\bm{\nabla}\phi)^2
        \label{eq:potential_1}\\
    \mu_\phi&= 2w\xi_p(\bm{\nabla}\psi\cdot\bm{\nabla}\phi+\psi{\nabla}^2\phi) .
        \label{eq:potential_2}        
\end{align}
In what follows we assume that the particles will interact with the interface hydrodynamically but not chemically and so we set $w=0$ in all the simulations reported below.
\subsubsection{Fluid dynamics}
The total velocity, which accounts for both the fluid and particle domains, is defined as
\begin{align}
        \bm{u}&=(1-\phi)\bm{u}_f+\phi\bm{u}_p,
\end{align}
where the first term gives the fluid velocity field, and the second term the particle velocity field. Then, the time evolution of this total flow field $\bm{u}$ is given by a modified version of the Navier-Stokes and continuity equations
\begin{align}
        \rho(\partial_t+\bm{u}\cdot\bm{\nabla})\bm{u} &= \bm{\nabla} \cdot \bm{\sigma}  + \rho(\phi\bm{f}_p+\bm{f}_{sq}) \label{eq:navier_stokes_model_h}\\
        &\quad - {\psi}\bm{\nabla} \mu_\psi - {\phi}\bm{\nabla} \mu_\phi ,\notag\\
        \bm{\nabla}\cdot\bm{u}&=0 \label{eq:divu}
\end{align}
where $\bm{\sigma}$ is the Newtonian stress tensor, defined in terms of the total fluid velocity as
\begin{align}
    \bm{\sigma} &= -p\bm{\mathsf{I}} + \eta[\bm{\nabla}\bm{u} +(\bm{\nabla}\bm{u})^T]
\end{align}
where $\eta$ is the spatially varying viscosity. The term $\phi\bm{f}_p$ appearing on the right-hand side of Eq.~\eqref{eq:navier_stokes_model_h} is the body force required to satisfy the rigidity constraint of the particles, the term $\phi\bm{f}_{sq}$ is the force needed to enforce the ``squirming'' boundary condition at their surface of the swimmers (Eq.~\eqref{eq:sq2}), and final two terms come from the binary-fluid nature of the host fluid.

In the present study, to keep the system as simple as possible, we assume that fluids $A$ and $B$ are strongly immiscible, but otherwise possess identical physical properties, except for their viscosity. Let $\eta_A$ and $\eta_B$ represent the viscosity of fluids A and B, respectively, and $\eta_p$ the viscosity of the particle domains. 
The total phase-dependent viscosity $\eta(\bm{r})$ is defined as
\begin{align}
        \eta(\bm{r}) &= \eta_A\psi_A(\bm{r})+\eta_B\psi_B(\bm{r})+\eta_p\phi(\bm{r})\label{vis}\\
        &= \eta_A\left(\psi_A(\bm{r}) + \lambda \psi_B(\bm{r})\right)+\eta_p\phi(\bm{r}),\notag
\end{align}
where  $\lambda=\eta_B/\eta_A$ is the fluid viscosity ratio. 

\section{Results}
\begin{figure}[htb!]
    \includegraphics[width=0.9\linewidth]{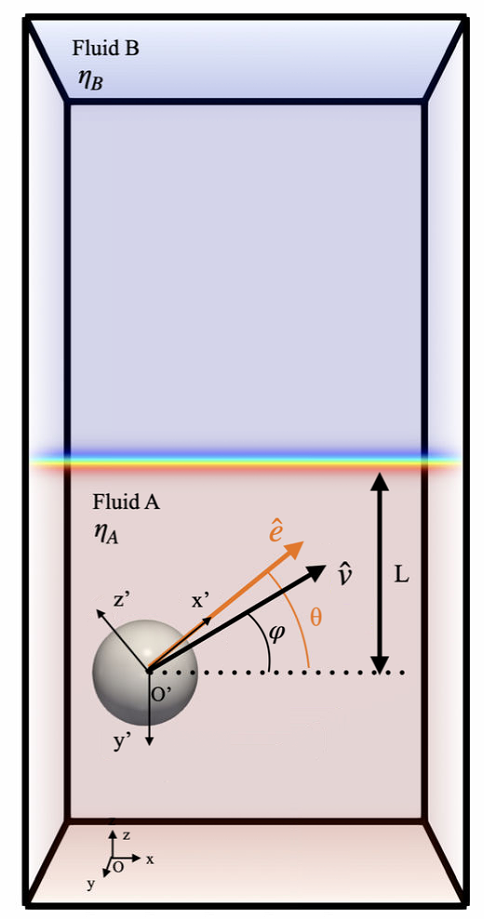}
\caption{\label{fig:schematic} Schematic representation of a single swimmer near an $A$/$B$ fluid-fluid interface normal to $\hat{z}$. The angles $\theta$ 
and $\varphi$ are here 
the angles between the the polar axis and the interface, and the direction of motion and the interface, respectively.}
\end{figure}

In this study, we conducted three dimensional direct numerical simulations (DNS) of a single particle near an $A$/$B$ fluid--fluid interface. 
In all cases, we use a rectangular simulation box of dimensions $(l_x,l_y,l_z) = (32\Delta, 32\Delta, 64\Delta)$ , where $\Delta$ is the grid spacing and the unit of length. Periodic boundary conditions are established in all directions. The radius of the squirmer is set to $a=4\Delta$. 
The parameter $B_1$ in Eq.~\eqref{eq:sq2} is set to $0.015$, corresponding to a single-particle steady-state velocity of $U_0=2/3B_1=0.01$. 
The mobility $\kappa$ (Eq.~\eqref{eq:cahn_hilliard}), and the mass densities for both fluids and particles $\rho=\rho_A=\rho_B=\rho_p$ are all set to $1$. 
We set the viscosity of the particle and fluid $A$ to be equal to unity, $\eta_p=\eta_A=1$, and vary the viscosity of fluid $B$ in the range $1/10 \le \eta_B\le 10$.
Then, the particle Reynolds number is $Re=\rho{U_0}a/\eta$ is $0.08$ in fluid A, while in fluid B it can range from $0.008$ to $0.8$.
The fluid--fluid interface thickness $\xi_{f}$ is of order unity with the present choice of parameter $\alpha=1$ in Eq.~\eqref{eq:freeenergy}, and the particle--fluid interface thickness $\xi_{p}$ is set to $2$.

A schematic representation of our system is given in Fig.~\ref{fig:schematic}, which shows a single swimmer near a fluid--fluid interface. The system is initialized to be phase separated along the $z$ direction. The swimmer is initially located in Fluid $A$ (the host fluid).
The distance between the center of mass of the swimmer and the interface is $L$, with $L_{t=0}=-16\Delta$ unless noted otherwise. 
This initial distance from the interface is large enough to allow the particle to attain its steady-state velocity before any appreciable particle/interface interactions are observed. 
The orientation of the interface is specified by its normal vector, $\uvec{z}$, that of the swimmer's motion by it normalized velocity vector $\uvec{v}=\bm{V}/\lvert\bm{V}\rvert$, which need not correspond to its polar axis $\uvec{e}$ taken to be parallel to the body frame $\uvec{x}'$-axis.
The initial orientation of the particle is fixed to lie in the $x-z$ plane, which will constrain it's motion to this plane (i.e., $V_y = 0$).
The x-component of the position is $R_x$.
The orientation angle $\theta=\arcsin{(\uvec{z}\cdot{\uvec{e}})}$ 
is defined as the angle between the the polar axis and the interface, while the direction of motion is defined by $\varphi = \arcsin{(\uvec{z}\cdot{\uvec{v}})}$.

\begin{figure*}
    \includegraphics[width=1\linewidth]{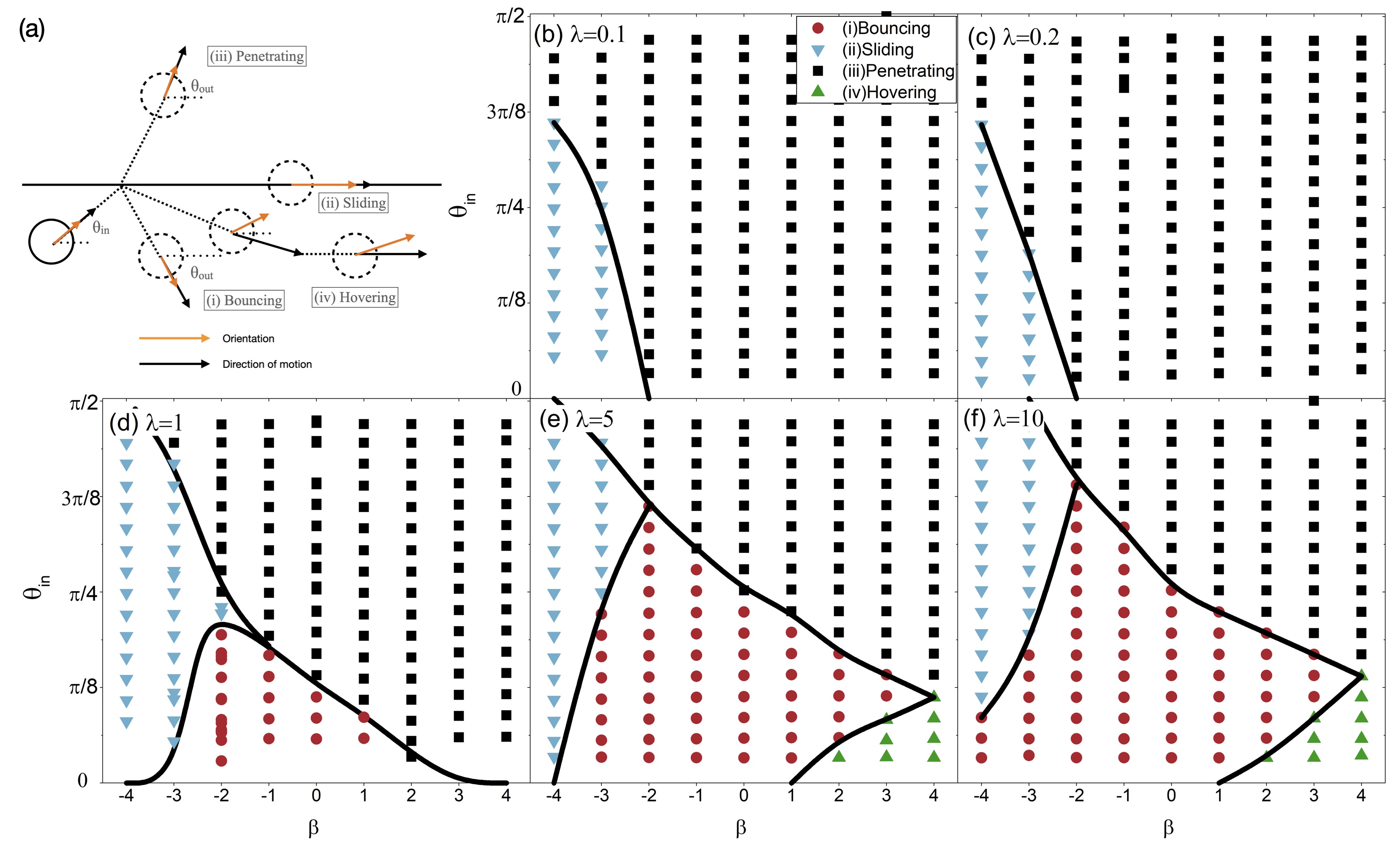}
\caption{\label{fig:phase} (a) Graphical illustration of the four collision modes of a swimmer with an interface. Black arrows indicate the swimmer's direction of motion, yellows arrows the swimmer's orientation; (b)--(f) Diagram showing how these modes depend on the initial incidence angle $\theta_{in}$ and swimmer type $\beta$ for various viscosity ratios (b) $\lambda=0.1$, (c) $\lambda=0.2$, (d) $\lambda=1$, (e) $\lambda=5$,  and (f)$\lambda=10$. 
}
\end{figure*}

To understand the dynamics of a swimmer near an interface for fluids with mismatched viscosities, we conduct a series of simulations in which the swimmer is initially in host fluid $A$, approaches the interface at an incoming angle $\theta_{in}$, and exits this ``collision'' with an outgoing angle $\theta_{out}$. Given the geometry of our setup, we can focus only on collisions with $\theta_{in}>0$ (i.e., the swimmer collides with the top interface), as those for $\theta_{in}<0$ are equivalent due to the reflection symmetry 
. We consider various initial angles $\theta_{in}$ and swimming parameters $\beta$, together with a variety of different viscosity ratios $\lambda=0.1$, $0.2$, $1$, $5$, and $10$, in order to construct a phase diagram for the four distinct dynamical modes that result: (i)``penetrating'', (ii)``sliding'', (iii)``bouncing'' and (iv)``hovering'', as illustrated in Fig. \ref{fig:phase}(a). The first three of these modes are also observed in the case of isoviscous fluids ($\lambda=1$), as reported in our previous work\cite{feng2022dynamics}. In case (i), the swimmer approaches the interface and exhibits a significant rotation within the interfacial domain, bouncing back into the host fluid $A$ and avoiding fluid $B$, leaving the interface with $\theta_{out}<0$. In case (ii), the swimmer becomes trapped at the interface, swimming in the $x-y$ plane with $\theta_{out}=0$. In case (iii), the swimmer crosses the interfacial region separating the fluids, swimming into fluid B with $\theta_{out}>0$.
In case (iv), which is only observed for $\lambda\ne 0$, the swimmer direction of motion shows characteristic oscillations as it approaches and turns away from the interface, before eventually swimming parallel to the interface at a fixed distance greater than the particle radius $|L|>a$. Unlike for the bouncing motion, where the swimmer is able to reorient and swim away from the interface with $\theta<0$, in the hovering motion the swimmer exhibits a partial reorientation to smaller angles, but it is always pointing towards the interface $\theta>0$. At the steady-state, the self-propulsion will balance with the hydrodynamic interactions with the interface, allowing the swimmer to propel itself parallel to the interface, even though the orientation of the swimmer is not aligned with it.
Our results are summarized in the phase diagram of Fig.~\ref{fig:phase}(b)--(f).

In our previous work on isoviscous systems, we have investigated how the collision dynamics depends on the angle of approach $\theta_{in}$ and the swimming mode $\beta$.  For weak swimmers, we observed either penetrating or bouncing motion, depending on the initial orientation: penetration (bouncing) for large (small) angle magnitudes. The bouncing motion is more prevalent for pushers ($\beta<0$) than pullers ($\beta > 0$), with the latter able to penetrate the interface at smaller angles. Strong pushers usually slide on the interface. The main role of the swimming mode $\beta$ is to shift the boundary between penetrating and bouncing domains. These results are summarized in the Fig.~\ref{fig:phase}(d).

In the case of mismatched viscosities, the focus of the current work, we found that the viscosity ratio $\lambda$ has a significant effect on the swimmer's motion, as can be seen in Fig.~\ref{fig:phase}.
For viscosity ratios less than unity $\lambda < 1$,  corresponding to swimmers starting in  the high viscosity fluid, the bouncing mode is never observed, rather the penetrating mode dominates for all but the strongest pushers. For such strong pushers, the sliding state can also be observed at small to moderate incoming angles.  Fig.~\ref{fig:phase}(b)-(c) shows the results for $\lambda=0.1$ and $0.2$. For viscosity ratios larger than unity $\lambda > 1$, corresponding to a swimmer starting in the lower viscosity fluid, the bouncing mode dominates. Thus, we infer that the the viscosity gradient will tend to propel swimmers towards regions of low viscosity and, the swimmers will exhibit a form of viscotaxis, with a preference for the low viscosity fluid. Finally, for strong pullers at small initial angles, a new dynamical ``hovering'' mode emerges, while  strong pushers still exhibit  sliding motion. These results are summarized in Fig.~\ref{fig:phase}(e)-(f), for $\lambda = 5$ and $10$.

\begin{figure*}
    \includegraphics[width=1\linewidth]{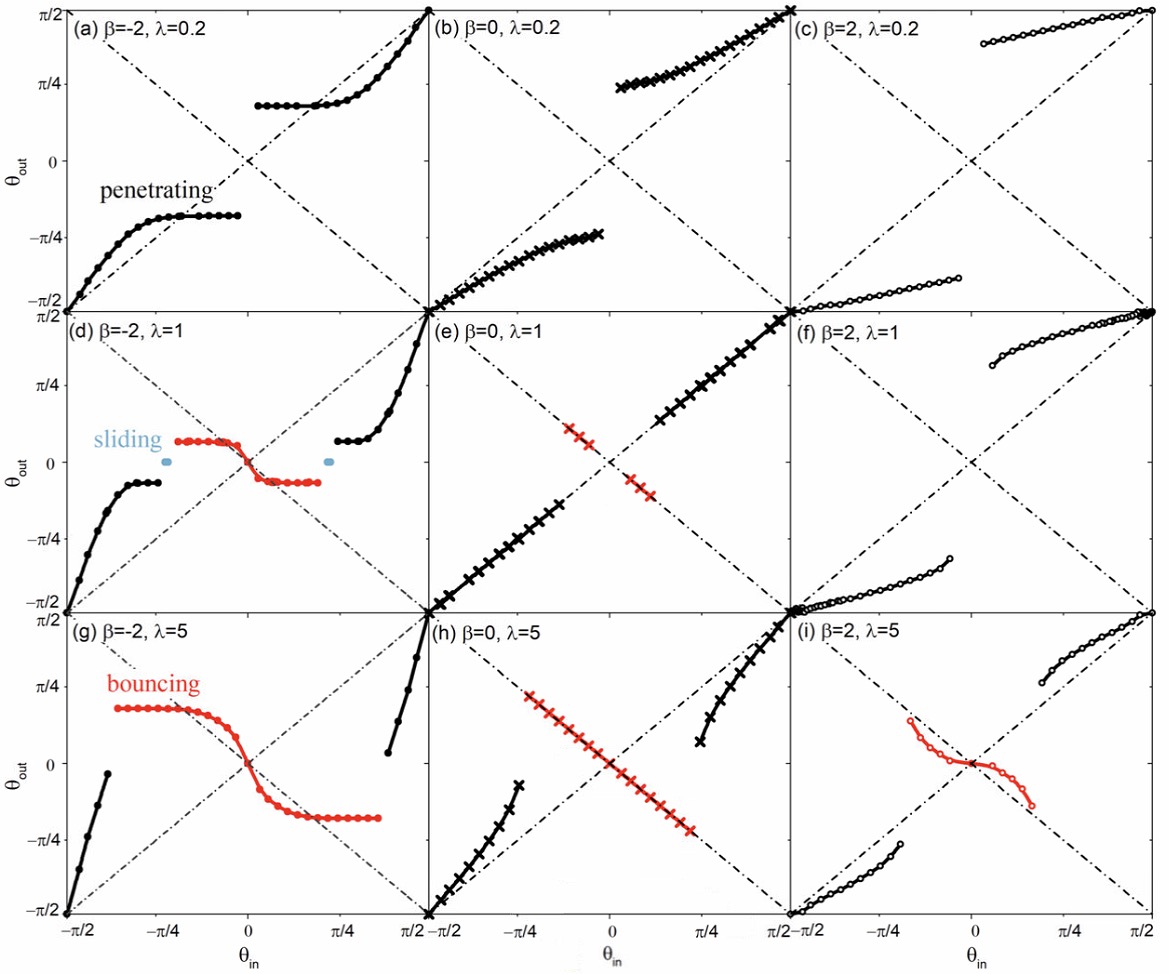}
\caption{\label{fig:change} Changes in the orientation angle $\theta$ for different viscosity ratios (a)--(c) $\lambda=0.2$; (d)--(f) $\lambda=1$; (g)--(i) $\lambda=5$ with swimmers with various swimming types $\beta=-2$, $0$ and $2$ after a single collision with the interface. This corresponds to a one-to-one mapping from $\theta_{in}$ to $\theta_{out}$. 
}
\end{figure*}

We further investigate the effect of the viscosity ratio $\lambda$ on the swimmers' collision dynamics, by computing the map $f$ relating the initial angle $\theta_{in}$ to the outgoing angle $\theta_{out}$, after a single collision with the interface. These collision maps are presented in Fig.~\ref{fig:change}, for $\beta=-2,0,2$ swimmers and viscosity ratios $\lambda=0.2, 1, 5$. Except for the bouncing motion of the neutral particles, which is absent for $\lambda=0.2$ and extends over a wider range of incoming angles for $\lambda=5$, the maps for $\lambda\ne 1$ exhibit clear deviations compared to the isoviscous results ($\lambda=1$). To illustrate this, consider the effect of $\lambda$ on the collision dynamics of pushers ($\beta=-2$). Comparing with the isoviscous system, shown in Fig.~\ref{fig:change}(d), the threshold angle $|\theta_c|$ dividing the bouncing and penetrating behaviors shifts to larger values when the swimmer is initially in the low-viscosity fluid (i.e., $\lambda=5$), as shown in Fig.~\ref{fig:change}(g). Furthermore, the outgoing angle $\theta_{out}$ 
for the penetrating/bouncing motion shows an overall decrease or increase depending on the viscosity ratio. This effect is particularly obvious for the penetrating motion, where the outgoing angle magnitude will be smaller than for the isoviscous case for the same initial angle $\theta_{in}$.
The opposite trends are observed for swimmers initially located in the higher viscosity fluid ($\lambda=0.2$), as shown in Fig.~\ref{fig:change}(a). 

\begin{figure*}[tbh]
    \includegraphics[width=1\linewidth]{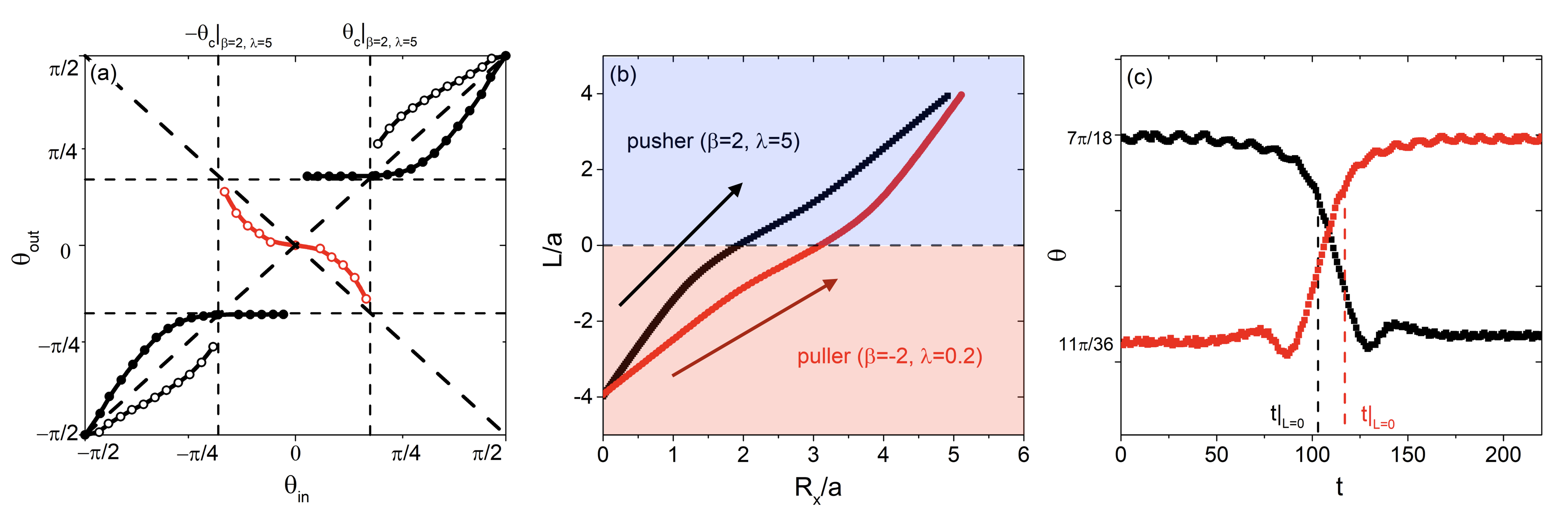}
\caption{\label{fig:sym_p} (a) Collisions maps, providing the change in the orientation angle $\theta$ for a puller ($\beta=2$) at $\lambda=5$ and a pusher ($\beta=-2$) at $\lambda=1/5$, marked by empty and filled circles, respectively. The black symbols represent the penetrating motion, red symbols the bouncing motion. (b) The particle trajectory and (c) the time evolution of the orientation angle $\theta$, for swimmers initially located at $L_{t=0}=-4a$. The pusher with $\beta=-2$, $\theta_{in}=7\pi/18$, and $\lambda=5$ is marked in black, while the puller with $\beta=2$, $\theta_{in}=11\pi/36$, and $\lambda=1/5$ is marked in red.}
\end{figure*}

Similar to the isoviscous case, where the pusher/puller duality is evident in the symmetry of the $\theta_{in}-\theta_{out}$ map and the particle trajectories at equal swimming strength $|\beta|$,  the results for mismatched viscocities also show a clear symmetry, despite the strong influence of $\lambda$ on the motion of the swimmer near the interface.
We first focus on the penetrating motion, because this was the mode that most clearly evidenced this symmetry for $\lambda=1$. For a single penetrating process, the trajectories and angular changes mirror each other. To illustrate this, we compare the map $f:\theta_{in} \to \theta_{out}$ for swimmers with $\beta=\pm2$, as shown in Fig.~\ref{fig:sym_p}(a). 
The mapping for the penetrating motion $f_p$ is symmetric about diagonal $\theta_{out}=\theta_{in}$. That is, the mapping for the pusher is the inverse of the mapping for the corresponding puller, under an inversion of the viscocity ratio, such that $f_p|_{\beta=-2, \lambda=1/5}=f_p^{-1}|_{\beta=2, \lambda=5}$. The threshold angle that divides the bouncing and penetrating motion for the pusher $\theta_c$ is equal to the maximum outgoing angle for the puller.
To illustrate this, we consider pushers ($\beta=-2$) with $\theta_{in}=7\pi/18$ at $\lambda=5$, and pullers ($\beta=2$) with $\theta_{in}=11\pi/36$ at $\lambda=0.2$ as representative examples.
Both swimmers are initially set at $L_{t=0}=-4a$. 
According to Fig.~\ref{fig:sym_p}(b), the trajectory of the puller (pusher) before it reaches the interface $L<0$, is same as the trajectory of pusher (puller) after it leaves the interface $L>0$.  
As expected, this symmetry is also evident in the evolution of the orientation angle, as shown in Fig.~\ref{fig:sym_p}(c).  This behaviour is seen for all swimmer types. In summary, the penetrating motion for pushers (pullers) at viscosity ratio $\lambda$, is the inverse of the penetrating motion for pullers (pushers) at viscosity ratio $1/\lambda$, such that $f_p|_{\beta, \lambda}=f_p^{-1}|_{-\beta, 1/\lambda}$. When $\lambda = 1$, we recover the results of our previous work, $f_p|_{\beta, \lambda=1}=f_p^{-1}|_{-\beta, \lambda=1}$\cite{feng2022dynamics}.

\begin{figure*}[tbh]
    \includegraphics[width=1\linewidth]{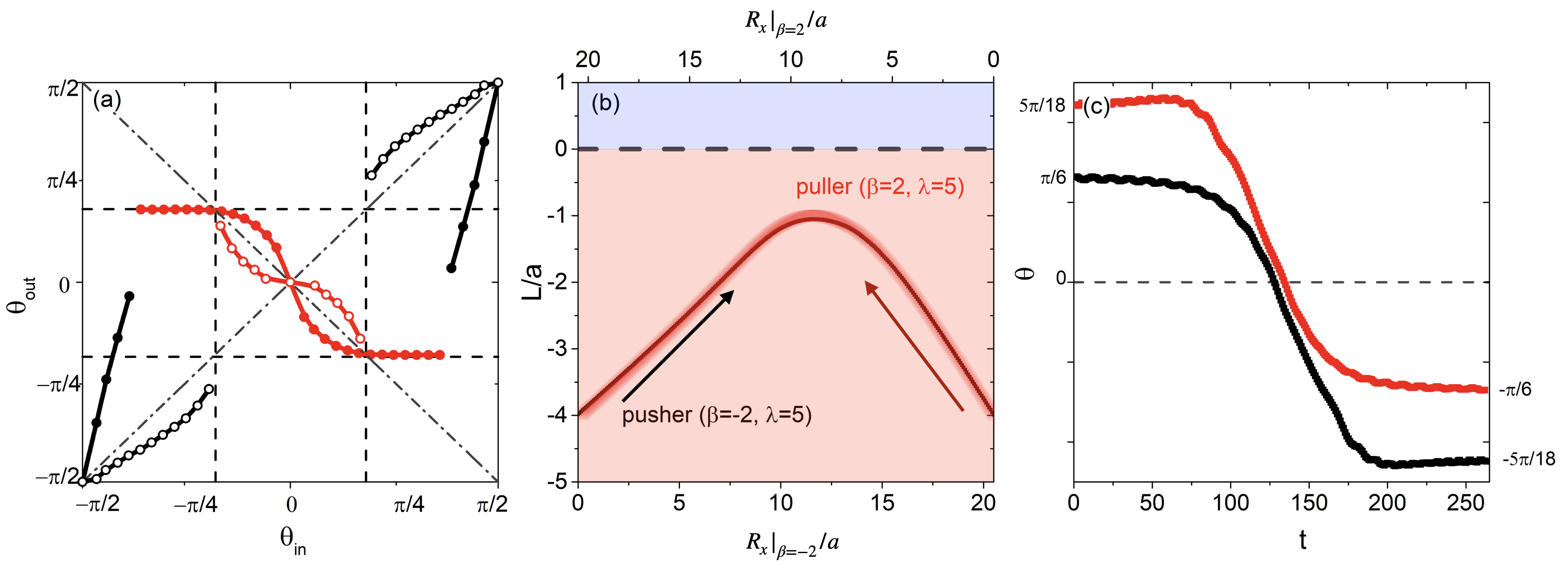}
\caption{\label{fig:sym_b} (a) Collision maps, providing the change in the orientation angle $\theta$ for a puller ($\beta=2$) and a pusher ($\beta=-2$) at $\lambda=5$, marked by empty and filled circles respectively; (b) the particle trajectory and (c) the time evolution of the orientation angle $\theta$ for swimmers initially located at $L_{t=0}=-4a$. The pusher ($\beta=-2$) with $\theta_{in}=\pi/6$ is marked in black, while the puller ($\beta=2$) with $\theta_{in}=5\pi/18$ is marked in red.
}
\end{figure*}

We now consider the bouncing motion of pushers and pullers. In view of the observations that a wide variety of swimmer orientations will result in penetration of the interface when the swimmers are moving into the low-viscosity fluid ($\lambda<1$) and that the pullers tend to penetrate at $\lambda=1$, we will focus on $\lambda>1$. 
According to Fig. \ref{fig:sym_b}(a), the mapping for the bouncing motion $f_b$ of swimmers $\beta=\pm2$ at a high viscosity interface ($\lambda=5$) is symmetric about the diagonal $\theta_{out}=-\theta_{in}$. Thus, the pusher/puller duality can be expressed as $f_b|_{\beta, \lambda}=-f_b^{-1}|_{-\beta, \lambda}$. This can be seen by comparing the results obtained from simulations of a single bouncing process, i.e. a pusher ($\beta=-2$) with $\theta_{in}=-\pi/6$ and a puller ($\beta=2$) with $\theta_{in}=-5\pi/18$, both under $\lambda=5$. The time-evolution of the particle positions and orientations for this process are shown in Fig.~\ref{fig:sym_b}(b-c). The pusher/puller trajectories are mirror images of each other, resulting from the time reversibility at low Reynolds number. 
That is, the trajectory for the bouncing motion of a pusher (puller) corresponds to the (time-reversed) bouncing motion of a puller (pusher).

\begin{figure*}[tbh]
    \includegraphics[width=1\linewidth]{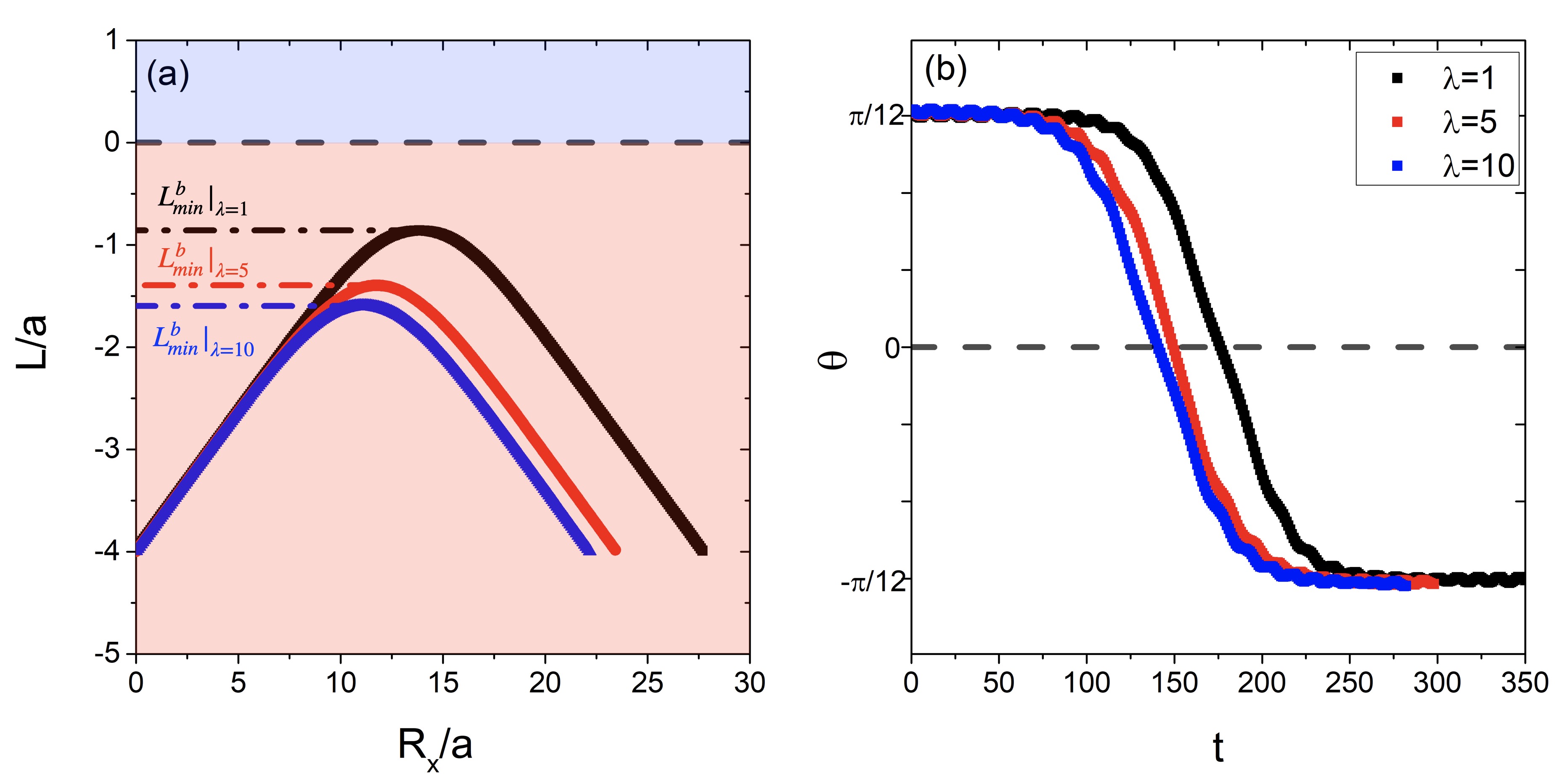}
\caption{\label{fig:sym_0} (a) The particle trajectory and (b) the time evolution of the orientation angle $\theta$, for neutral particles ($\beta=0$) at $\lambda=1$, $5$ and $10$. All swimmers are initially located at $L_{t=0}=-4a$, with orientation $\theta_{in}=\pi/12$.}
\end{figure*}

The bouncing motion for neutral particles, corresponding to the red curves in Fig.~\ref{fig:change}(e) and (h), is particularly interesting in how insensitive it is to $\lambda$, in contrast to pusher/pullers. Even though the threshold angle $\theta_c$ is seen to increase upon increasing $\lambda$
the outgoing angle $\theta_{out}$ is always equal in magnitude to the approaching angle $\theta_{in}$, regardless of the viscosity ratio $\lambda$. This represents a special case of the bouncing map obtained for pushers/pullers, $f_b|_{\beta, \lambda}=-f_b^{-1}|_{-\beta, \lambda}$, with $\beta=0$. 
In order to further validate this observation, we conducted a series of simulations for neutral particles with various fluid viscosity ratios $\lambda=0.1$, $0.2$, and $1$.
The particles initially approach the interface with $\theta_{in}=\pi/12$. 
The effect of the viscosity is evident in the distance of shortest approach to the interface $|L^{b}_{min}|$, which decreases with decreasing $\lambda$, as shown in Fig.~\ref{fig:sym_0}(a).
This indicates that the particle is sensitive to the viscosity gradient, and that a larger change in viscosity results in longer-range interactions, which allow the particle to start its reorientation process to escape the interface earlier~\ref{fig:sym_0}(b).

\begin{figure*}[tbh]
    \includegraphics[width=1\linewidth]{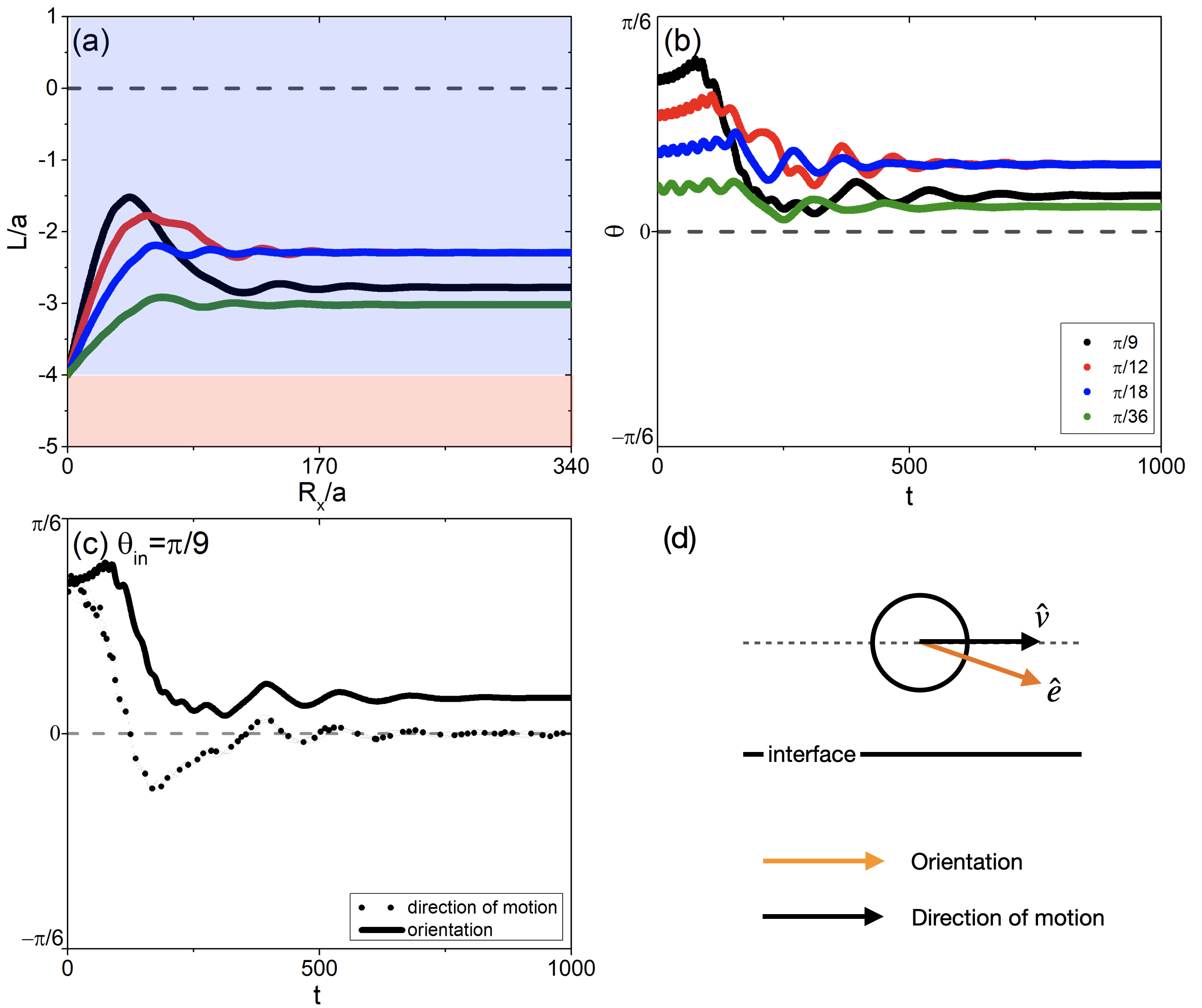}
\caption{\label{fig:hovering} The new dynamical mode of ``hovering motion results in particles swimming in the low viscosity fluid parallel to the interface without touching it. (a) The particle trajectories are shown as well as (b) the time evolution of the orientation angle $\theta$ for pullers ($\beta=4$) at $\lambda=10$ for various approach orientations $\theta_{in}$, with the swimmers initially located at $L_{t=0}=-4a$;
(c) Shows the time evolution for $\theta$ (solid line) and $\varphi$ (dotted line), i.e., the orientation and direction of motion, respectively, for a puller ($\beta=4$) at $\lambda=10$, with $\theta_{in}=\pi/9$. 
(d) Graphical illustration of the swimming state for a strong puller in a typical hovering motion.}
\end{figure*}

Finally, for strong pullers, we have observed a new ``hovering'' mode, different from the penetrating, bouncing, or sliding motions. To study this motion, we conducted simulations of for puller with $\beta=4$, at various approaching angles $\theta_{in}=\pi/9$, $\pi/12$, $\pi/18$ and $\pi/36$, with all particles initially located at $L_{t=0}=-4a$. The swimmer approaches the interface, initially turning towards it, while showing oscillations in its direction of motion. The swimmer then partially turns away from the interface, with increased oscillations in its orientation, before settling into the hovering motion, moving parallel to the interface at a fixed distance (see Fig.~\ref{fig:hovering}(a)).
We note that the orientation of the swimmer is not completely aligned with the interface, it is always pointing towards the interface, with a positive $\theta$ throughout the entire process, as shown in Fig. \ref{fig:hovering}(b).
Taking the dynamics of a $\beta=4$ puller, with initial angle $\theta_{in}=\pi/9$, for $\lambda=10$ as a reference (see Fig. \ref{fig:hovering}(c)), we can clearly see that the orientation of the swimmer and its direction of motion are not parallel.
In contrast to these pullers, strong pushers will show a sliding motion, in which they are adsorbed to and swim along the interface\cite{feng2022dynamics}.

\section{Discussion}

We have studied the effect of a viscosity ratio in a binary phase-separated fluid on the dynamics of swimmers near the interface.  We notice that swimmers show a preference towards the lower viscosity fluid.
If one considers an ensemble of swimmers, with a variety of incidence angles, the effect of the viscosity gradient will be to propel swimmers towards regions of low viscosity. This will tend to result in an enrichment of swimmers in the low viscosity fluid. This type of viscotaxis is consistent with previous theoretical investigations on swimmers in viscosity gradients.\cite{datt2019active,liebchen2018viscotaxis,dandekar2020swimming,eastham2020axisymmetric,gidituri2022reorientation} These studies showed that a squirmer in a weak viscosity gradient will  reorient in the direction of the lower viscosity region (negative viscotaxis), regardless of the swimming mode. This effect has been confirmed in experiments, such as the puller-like alga Chlamydomonas reinhardtii, which was observed to accumulate in low-viscosity zones at sufficiently strong gradients. \cite{daniels1980aspects,takabe2017viscosity,coppola2021green,stehnach2021viscophobic,lopez2021dynamics}

We observe a new mode of motion for strong pullers which we refer to as hovering. This mode was not observed for swimmers in fluids of equal viscosity. A similar mode as been reported in previous work that focused on pullers near a wall, rather than a deformable fluid-fluid interface at finite $\lambda$ \cite{ishimoto2013squirmer,lintuvuori2016hydrodynamic,li2014hydrodynamic,pagonabarraga2013structure,shen2018hydrodynamic}. Although it is difficult to make a direct comparison with our work, the dynamics previously reported for pullers near rigid walls is similar to the hovering motion we observe. 
In both their work and ours, strong pullers first exhibit oscillating motion before eventually swimming parallel to the interface at a constant separation, maintaining an orientation tilted towards the interface (i.e., the particle orientation is not aligned with the direction of motion). 
Some understanding of this similarity can be obtained by noting that a solid wall can be regarded as a fluid-fluid interface in the limit $\lambda\to\infty$. Furthermore, for the hovering motion, the swimmers remain in the host (low viscosity) fluid, with the distance of closest approach increasing as the viscosity ratio $\lambda$ increases. Thus, the interaction with the interface can be expected to be well represented by the far-field approximations used in previous studies\cite{li2014hydrodynamic}. Finally, in order to rule out the possibility that the observed dynamics is due to inertial effects, we also conducted simulations with a smaller Reynolds number by decreasing the velocity of swimmers. We obtained equivalent trajectories, which indicates that the inertial effects are negligible in our present simulations.

\section{Conclusions}

In this work, we have analyzed the dynamics of swimmers approaching a fluid-fluid interface between phase-separated fluids with distinct viscosities. 
The viscosity ratio $\lambda$, together with the swimming mode $\beta$ (pusher or puller), determine the outcome of collisions of the swimmer with the interface. Such collisions are shown to give rise to four distinct dynamics modes: bouncing, sliding, penetrating and hovering. 
Compared with the results obtained for isoviscous systems $\lambda=1$, we observe that the swimmer exhibits a preference towards the lower viscosity fluid (i.e. viscotaxis). This preference should be understood in the sense that, for a wide distribution of contact angles, more swimmers will transition into the low viscous environment than vice-versa.  This means that a typical swimmer, drawn from this distribution, that starts in a low viscosity fluid is more likely to bounce back (reflect) at the interface, while a swimmer starting in the high viscosity fluid is more likely to penetrate the interface and swim into the lower viscosity fluid. 
Even though the outgoing angle $\theta_{out}$ for the penetrating/bouncing motion is mainly determined by the swimming type $\beta$ and the initial angle $\theta_{in}$, the precise functional dependence depends on the viscosity ratio $\lambda$.
A duality between pushers/pullers can be clearly seen in the trajectories for penetrating and bouncing swimmers. In addition, we observed that strong pullers, initially located in the lower viscosity fluid, can exhibit a new type of hovering motion, moving parallel to the interface at a fixed distance $L/a > 1$, which is not observed for isoviscous systems.

Our study represents a detailed analysis of the role played by the viscosity ratio on the dynamics of swimmers near an interface. This improves our understanding of swimming in complex fluids environments, and may help with the interpretion of similar physiological and biological phenomena. 
Further developments of this work may allow a numerical investigation of more complex interfacial structures, such as curved interfaces, e.g., swimming near a spherical or tubular domain. 

\begin{acknowledgments}
The authors express their gratitude to Dr. Hiroto Ozaki and Dr. Takeshi Aoyagi for their collaboration and outstanding contributions to the simulation software development.
R.Y. acknowledges helpful discussions with Profs. Hajime Tanaka and Akira Furukawa.
This work was supported by the Grants-in-Aid for Scientific Research (JSPS KAKENHI) under grant nos. JP 20H00129, 20H05619, and 20K03786 and by the NEDO Project (JPNP16010).
This work was supported by JST, the establishment of university fellowships towards the creation of science technology innovation, Grant Number JPMJFS2123.

\end{acknowledgments}

\section*{Data Availability}
The data that support the findings of this study are available from the corresponding author upon reasonable request.

\appendix
\section{Software}
All simulations presented in this paper were conducted using the open-source version of the KAPSEL DNS software.
KAPSEL has been developed in our lab to simulate the dynamics of solid particles dispersed in complex fluids.
Detailed descriptions of KAPSEL are available online (https://kapsel-dns.com).

\bibliography{ref.bib}

\begin{thebibliography}{40}%
\makeatletter
\providecommand \@ifxundefined [1]{%
 \@ifx{#1\undefined}
}%
\providecommand \@ifnum [1]{%
 \ifnum #1\expandafter \@firstoftwo
 \else \expandafter \@secondoftwo
 \fi
}%
\providecommand \@ifx [1]{%
 \ifx #1\expandafter \@firstoftwo
 \else \expandafter \@secondoftwo
 \fi
}%
\providecommand \natexlab [1]{#1}%
\providecommand \enquote  [1]{``#1''}%
\providecommand \bibnamefont  [1]{#1}%
\providecommand \bibfnamefont [1]{#1}%
\providecommand \citenamefont [1]{#1}%
\providecommand \href@noop [0]{\@secondoftwo}%
\providecommand \href [0]{\begingroup \@sanitize@url \@href}%
\providecommand \@href[1]{\@@startlink{#1}\@@href}%
\providecommand \@@href[1]{\endgroup#1\@@endlink}%
\providecommand \@sanitize@url [0]{\catcode `\\12\catcode `\$12\catcode
  `\&12\catcode `\#12\catcode `\^12\catcode `\_12\catcode `\%12\relax}%
\providecommand \@@startlink[1]{}%
\providecommand \@@endlink[0]{}%
\providecommand \url  [0]{\begingroup\@sanitize@url \@url }%
\providecommand \@url [1]{\endgroup\@href {#1}{\urlprefix }}%
\providecommand \urlprefix  [0]{URL }%
\providecommand \Eprint [0]{\href }%
\providecommand \doibase [0]{http://dx.doi.org/}%
\providecommand \selectlanguage [0]{\@gobble}%
\providecommand \bibinfo  [0]{\@secondoftwo}%
\providecommand \bibfield  [0]{\@secondoftwo}%
\providecommand \translation [1]{[#1]}%
\providecommand \BibitemOpen [0]{}%
\providecommand \bibitemStop [0]{}%
\providecommand \bibitemNoStop [0]{.\EOS\space}%
\providecommand \EOS [0]{\spacefactor3000\relax}%
\providecommand \BibitemShut  [1]{\csname bibitem#1\endcsname}%
\let\auto@bib@innerbib\@empty
\bibitem [{\citenamefont {Kirkman-Brown}\ and\ \citenamefont
  {Smith}(2011)}]{kirkman2011sperm}%
  \BibitemOpen
  \bibfield  {author} {\bibinfo {author} {\bibfnamefont {J.~C.}\ \bibnamefont
  {Kirkman-Brown}}\ and\ \bibinfo {author} {\bibfnamefont {D.~J.}\ \bibnamefont
  {Smith}},\ }\bibfield  {title} {\enquote {\bibinfo {title} {Sperm motility:
  is viscosity fundamental to progress?}}\ }\href@noop {} {\bibfield  {journal}
  {\bibinfo  {journal} {MHR: Basic science of reproductive medicine}\ }\textbf
  {\bibinfo {volume} {17}},\ \bibinfo {pages} {539--544} (\bibinfo {year}
  {2011})}\BibitemShut {NoStop}%
\bibitem [{\citenamefont {Gad{\^e}lha}\ \emph {et~al.}(2010)\citenamefont
  {Gad{\^e}lha}, \citenamefont {Gaffney}, \citenamefont {Smith},\ and\
  \citenamefont {Kirkman-Brown}}]{gadelha2010nonlinear}%
  \BibitemOpen
  \bibfield  {author} {\bibinfo {author} {\bibfnamefont {H.}~\bibnamefont
  {Gad{\^e}lha}}, \bibinfo {author} {\bibfnamefont {E.}~\bibnamefont
  {Gaffney}}, \bibinfo {author} {\bibfnamefont {D.}~\bibnamefont {Smith}}, \
  and\ \bibinfo {author} {\bibfnamefont {J.}~\bibnamefont {Kirkman-Brown}},\
  }\bibfield  {title} {\enquote {\bibinfo {title} {Nonlinear instability in
  flagellar dynamics: a novel modulation mechanism in sperm migration?}}\
  }\href@noop {} {\bibfield  {journal} {\bibinfo  {journal} {Journal of The
  Royal Society Interface}\ }\textbf {\bibinfo {volume} {7}},\ \bibinfo {pages}
  {1689--1697} (\bibinfo {year} {2010})}\BibitemShut {NoStop}%
\bibitem [{\citenamefont {Bj{\"o}rndahl}(2010)}]{bjorndahl2010usefulness}%
  \BibitemOpen
  \bibfield  {author} {\bibinfo {author} {\bibfnamefont {L.}~\bibnamefont
  {Bj{\"o}rndahl}},\ }\bibfield  {title} {\enquote {\bibinfo {title} {The
  usefulness and significance of assessing rapidly progressive spermatozoa},}\
  }\href@noop {} {\bibfield  {journal} {\bibinfo  {journal} {Asian Journal of
  Andrology}\ }\textbf {\bibinfo {volume} {12}},\ \bibinfo {pages} {33}
  (\bibinfo {year} {2010})}\BibitemShut {NoStop}%
\bibitem [{\citenamefont {Nguyen}\ and\ \citenamefont
  {Faivre}(2020)}]{nguyen2020targeted}%
  \BibitemOpen
  \bibfield  {author} {\bibinfo {author} {\bibfnamefont {H.~V.}\ \bibnamefont
  {Nguyen}}\ and\ \bibinfo {author} {\bibfnamefont {V.}~\bibnamefont
  {Faivre}},\ }\bibfield  {title} {\enquote {\bibinfo {title} {Targeted drug
  delivery therapies inspired by natural taxes},}\ }\href@noop {} {\bibfield
  {journal} {\bibinfo  {journal} {Journal of Controlled Release}\ }\textbf
  {\bibinfo {volume} {322}},\ \bibinfo {pages} {439--456} (\bibinfo {year}
  {2020})}\BibitemShut {NoStop}%
\bibitem [{\citenamefont {Ceylan}\ \emph {et~al.}(2018)\citenamefont {Ceylan},
  \citenamefont {Yasa}, \citenamefont {Yasa}, \citenamefont {Tabak},
  \citenamefont {Giltinan},\ and\ \citenamefont {Sitti}}]{ceylan20183d}%
  \BibitemOpen
  \bibfield  {author} {\bibinfo {author} {\bibfnamefont {H.}~\bibnamefont
  {Ceylan}}, \bibinfo {author} {\bibfnamefont {I.~C.}\ \bibnamefont {Yasa}},
  \bibinfo {author} {\bibfnamefont {O.}~\bibnamefont {Yasa}}, \bibinfo {author}
  {\bibfnamefont {A.~F.}\ \bibnamefont {Tabak}}, \bibinfo {author}
  {\bibfnamefont {J.}~\bibnamefont {Giltinan}}, \ and\ \bibinfo {author}
  {\bibfnamefont {M.}~\bibnamefont {Sitti}},\ }\bibfield  {title} {\enquote
  {\bibinfo {title} {3d-printed biodegradable microswimmer for drug delivery
  and targeted cell labeling},}\ }\href@noop {} {\bibfield  {journal} {\bibinfo
   {journal} {BioRxiv}\ ,\ \bibinfo {pages} {379024}} (\bibinfo {year}
  {2018})}\BibitemShut {NoStop}%
\bibitem [{\citenamefont {Mostaghaci}\ \emph {et~al.}(2017)\citenamefont
  {Mostaghaci}, \citenamefont {Yasa}, \citenamefont {Zhuang},\ and\
  \citenamefont {Sitti}}]{mostaghaci2017bioadhesive}%
  \BibitemOpen
  \bibfield  {author} {\bibinfo {author} {\bibfnamefont {B.}~\bibnamefont
  {Mostaghaci}}, \bibinfo {author} {\bibfnamefont {O.}~\bibnamefont {Yasa}},
  \bibinfo {author} {\bibfnamefont {J.}~\bibnamefont {Zhuang}}, \ and\ \bibinfo
  {author} {\bibfnamefont {M.}~\bibnamefont {Sitti}},\ }\bibfield  {title}
  {\enquote {\bibinfo {title} {Bioadhesive bacterial microswimmers for targeted
  drug delivery in the urinary and gastrointestinal tracts},}\ }\href@noop {}
  {\bibfield  {journal} {\bibinfo  {journal} {Advanced Science}\ }\textbf
  {\bibinfo {volume} {4}},\ \bibinfo {pages} {1700058} (\bibinfo {year}
  {2017})}\BibitemShut {NoStop}%
\bibitem [{\citenamefont {Park}\ \emph {et~al.}(2017)\citenamefont {Park},
  \citenamefont {Zhuang}, \citenamefont {Yasa},\ and\ \citenamefont
  {Sitti}}]{park2017multifunctional}%
  \BibitemOpen
  \bibfield  {author} {\bibinfo {author} {\bibfnamefont {B.-W.}\ \bibnamefont
  {Park}}, \bibinfo {author} {\bibfnamefont {J.}~\bibnamefont {Zhuang}},
  \bibinfo {author} {\bibfnamefont {O.}~\bibnamefont {Yasa}}, \ and\ \bibinfo
  {author} {\bibfnamefont {M.}~\bibnamefont {Sitti}},\ }\bibfield  {title}
  {\enquote {\bibinfo {title} {Multifunctional bacteria-driven microswimmers
  for targeted active drug delivery},}\ }\href@noop {} {\bibfield  {journal}
  {\bibinfo  {journal} {ACS nano}\ }\textbf {\bibinfo {volume} {11}},\ \bibinfo
  {pages} {8910--8923} (\bibinfo {year} {2017})}\BibitemShut {NoStop}%
\bibitem [{\citenamefont {Yan}\ \emph {et~al.}(2015)\citenamefont {Yan},
  \citenamefont {Zhou}, \citenamefont {Yu}, \citenamefont {Xu}, \citenamefont
  {Deng}, \citenamefont {Tang}, \citenamefont {Feng}, \citenamefont {Bian},
  \citenamefont {Zhang}, \citenamefont {Ferreira} \emph
  {et~al.}}]{yan2015magnetite}%
  \BibitemOpen
  \bibfield  {author} {\bibinfo {author} {\bibfnamefont {X.}~\bibnamefont
  {Yan}}, \bibinfo {author} {\bibfnamefont {Q.}~\bibnamefont {Zhou}}, \bibinfo
  {author} {\bibfnamefont {J.}~\bibnamefont {Yu}}, \bibinfo {author}
  {\bibfnamefont {T.}~\bibnamefont {Xu}}, \bibinfo {author} {\bibfnamefont
  {Y.}~\bibnamefont {Deng}}, \bibinfo {author} {\bibfnamefont {T.}~\bibnamefont
  {Tang}}, \bibinfo {author} {\bibfnamefont {Q.}~\bibnamefont {Feng}}, \bibinfo
  {author} {\bibfnamefont {L.}~\bibnamefont {Bian}}, \bibinfo {author}
  {\bibfnamefont {Y.}~\bibnamefont {Zhang}}, \bibinfo {author} {\bibfnamefont
  {A.}~\bibnamefont {Ferreira}},  \emph {et~al.},\ }\bibfield  {title}
  {\enquote {\bibinfo {title} {Magnetite nanostructured porous hollow helical
  microswimmers for targeted delivery},}\ }\href@noop {} {\bibfield  {journal}
  {\bibinfo  {journal} {Advanced Functional Materials}\ }\textbf {\bibinfo
  {volume} {25}},\ \bibinfo {pages} {5333--5342} (\bibinfo {year}
  {2015})}\BibitemShut {NoStop}%
\bibitem [{\citenamefont {Laage}\ and\ \citenamefont
  {Hynes}(2006)}]{laage2006molecular}%
  \BibitemOpen
  \bibfield  {author} {\bibinfo {author} {\bibfnamefont {D.}~\bibnamefont
  {Laage}}\ and\ \bibinfo {author} {\bibfnamefont {J.~T.}\ \bibnamefont
  {Hynes}},\ }\bibfield  {title} {\enquote {\bibinfo {title} {A molecular jump
  mechanism of water reorientation},}\ }\href@noop {} {\bibfield  {journal}
  {\bibinfo  {journal} {Science}\ }\textbf {\bibinfo {volume} {311}},\ \bibinfo
  {pages} {832--835} (\bibinfo {year} {2006})}\BibitemShut {NoStop}%
\bibitem [{\citenamefont {Ishimoto}\ and\ \citenamefont
  {Gaffney}(2013)}]{ishimoto2013squirmer}%
  \BibitemOpen
  \bibfield  {author} {\bibinfo {author} {\bibfnamefont {K.}~\bibnamefont
  {Ishimoto}}\ and\ \bibinfo {author} {\bibfnamefont {E.~A.}\ \bibnamefont
  {Gaffney}},\ }\bibfield  {title} {\enquote {\bibinfo {title} {Squirmer
  dynamics near a boundary},}\ }\href@noop {} {\bibfield  {journal} {\bibinfo
  {journal} {Physical Review E}\ }\textbf {\bibinfo {volume} {88}},\ \bibinfo
  {pages} {062702} (\bibinfo {year} {2013})}\BibitemShut {NoStop}%
\bibitem [{\citenamefont {Lintuvuori}\ \emph {et~al.}(2016)\citenamefont
  {Lintuvuori}, \citenamefont {Brown}, \citenamefont {Stratford},\ and\
  \citenamefont {Marenduzzo}}]{lintuvuori2016hydrodynamic}%
  \BibitemOpen
  \bibfield  {author} {\bibinfo {author} {\bibfnamefont {J.~S.}\ \bibnamefont
  {Lintuvuori}}, \bibinfo {author} {\bibfnamefont {A.~T.}\ \bibnamefont
  {Brown}}, \bibinfo {author} {\bibfnamefont {K.}~\bibnamefont {Stratford}}, \
  and\ \bibinfo {author} {\bibfnamefont {D.}~\bibnamefont {Marenduzzo}},\
  }\bibfield  {title} {\enquote {\bibinfo {title} {Hydrodynamic oscillations
  and variable swimming speed in squirmers close to repulsive walls},}\
  }\href@noop {} {\bibfield  {journal} {\bibinfo  {journal} {Soft Matter}\
  }\textbf {\bibinfo {volume} {12}},\ \bibinfo {pages} {7959--7968} (\bibinfo
  {year} {2016})}\BibitemShut {NoStop}%
\bibitem [{\citenamefont {Li}\ and\ \citenamefont
  {Ardekani}(2014)}]{li2014hydrodynamic}%
  \BibitemOpen
  \bibfield  {author} {\bibinfo {author} {\bibfnamefont {G.-J.}\ \bibnamefont
  {Li}}\ and\ \bibinfo {author} {\bibfnamefont {A.~M.}\ \bibnamefont
  {Ardekani}},\ }\bibfield  {title} {\enquote {\bibinfo {title} {Hydrodynamic
  interaction of microswimmers near a wall},}\ }\href@noop {} {\bibfield
  {journal} {\bibinfo  {journal} {Physical Review E}\ }\textbf {\bibinfo
  {volume} {90}},\ \bibinfo {pages} {013010} (\bibinfo {year}
  {2014})}\BibitemShut {NoStop}%
\bibitem [{\citenamefont {Pagonabarraga}\ and\ \citenamefont
  {Llopis}(2013)}]{pagonabarraga2013structure}%
  \BibitemOpen
  \bibfield  {author} {\bibinfo {author} {\bibfnamefont {I.}~\bibnamefont
  {Pagonabarraga}}\ and\ \bibinfo {author} {\bibfnamefont {I.}~\bibnamefont
  {Llopis}},\ }\bibfield  {title} {\enquote {\bibinfo {title} {The structure
  and rheology of sheared model swimmer suspensions},}\ }\href@noop {}
  {\bibfield  {journal} {\bibinfo  {journal} {Soft Matter}\ }\textbf {\bibinfo
  {volume} {9}},\ \bibinfo {pages} {7174--7184} (\bibinfo {year}
  {2013})}\BibitemShut {NoStop}%
\bibitem [{\citenamefont {Shen}, \citenamefont {W{\"u}rger},\ and\
  \citenamefont {Lintuvuori}(2018)}]{shen2018hydrodynamic}%
  \BibitemOpen
  \bibfield  {author} {\bibinfo {author} {\bibfnamefont {Z.}~\bibnamefont
  {Shen}}, \bibinfo {author} {\bibfnamefont {A.}~\bibnamefont {W{\"u}rger}}, \
  and\ \bibinfo {author} {\bibfnamefont {J.~S.}\ \bibnamefont {Lintuvuori}},\
  }\bibfield  {title} {\enquote {\bibinfo {title} {Hydrodynamic interaction of
  a self-propelling particle with a wall},}\ }\href@noop {} {\bibfield
  {journal} {\bibinfo  {journal} {The European Physical Journal E}\ }\textbf
  {\bibinfo {volume} {41}},\ \bibinfo {pages} {1--9} (\bibinfo {year}
  {2018})}\BibitemShut {NoStop}%
\bibitem [{\citenamefont {Volpe}\ \emph {et~al.}(2011)\citenamefont {Volpe},
  \citenamefont {Buttinoni}, \citenamefont {Vogt}, \citenamefont
  {K{\"u}mmerer},\ and\ \citenamefont {Bechinger}}]{volpe2011microswimmers}%
  \BibitemOpen
  \bibfield  {author} {\bibinfo {author} {\bibfnamefont {G.}~\bibnamefont
  {Volpe}}, \bibinfo {author} {\bibfnamefont {I.}~\bibnamefont {Buttinoni}},
  \bibinfo {author} {\bibfnamefont {D.}~\bibnamefont {Vogt}}, \bibinfo {author}
  {\bibfnamefont {H.-J.}\ \bibnamefont {K{\"u}mmerer}}, \ and\ \bibinfo
  {author} {\bibfnamefont {C.}~\bibnamefont {Bechinger}},\ }\bibfield  {title}
  {\enquote {\bibinfo {title} {Microswimmers in patterned environments},}\
  }\href@noop {} {\bibfield  {journal} {\bibinfo  {journal} {Soft Matter}\
  }\textbf {\bibinfo {volume} {7}},\ \bibinfo {pages} {8810--8815} (\bibinfo
  {year} {2011})}\BibitemShut {NoStop}%
\bibitem [{\citenamefont {Fadda}, \citenamefont {Molina},\ and\ \citenamefont
  {Yamamoto}(2020)}]{fadda2020dynamics}%
  \BibitemOpen
  \bibfield  {author} {\bibinfo {author} {\bibfnamefont {F.}~\bibnamefont
  {Fadda}}, \bibinfo {author} {\bibfnamefont {J.~J.}\ \bibnamefont {Molina}}, \
  and\ \bibinfo {author} {\bibfnamefont {R.}~\bibnamefont {Yamamoto}},\
  }\bibfield  {title} {\enquote {\bibinfo {title} {Dynamics of a chiral swimmer
  sedimenting on a flat plate},}\ }\href@noop {} {\bibfield  {journal}
  {\bibinfo  {journal} {Physical Review E}\ }\textbf {\bibinfo {volume}
  {101}},\ \bibinfo {pages} {052608} (\bibinfo {year} {2020})}\BibitemShut
  {NoStop}%
\bibitem [{\citenamefont {Lauga}\ \emph {et~al.}(2006)\citenamefont {Lauga},
  \citenamefont {DiLuzio}, \citenamefont {Whitesides},\ and\ \citenamefont
  {Stone}}]{lauga2006swimming}%
  \BibitemOpen
  \bibfield  {author} {\bibinfo {author} {\bibfnamefont {E.}~\bibnamefont
  {Lauga}}, \bibinfo {author} {\bibfnamefont {W.~R.}\ \bibnamefont {DiLuzio}},
  \bibinfo {author} {\bibfnamefont {G.~M.}\ \bibnamefont {Whitesides}}, \ and\
  \bibinfo {author} {\bibfnamefont {H.~A.}\ \bibnamefont {Stone}},\ }\bibfield
  {title} {\enquote {\bibinfo {title} {Swimming in circles: motion of bacteria
  near solid boundaries},}\ }\href@noop {} {\bibfield  {journal} {\bibinfo
  {journal} {Biophysical journal}\ }\textbf {\bibinfo {volume} {90}},\ \bibinfo
  {pages} {400--412} (\bibinfo {year} {2006})}\BibitemShut {NoStop}%
\bibitem [{\citenamefont {Di~Leonardo}\ \emph {et~al.}(2011)\citenamefont
  {Di~Leonardo}, \citenamefont {Dell’Arciprete}, \citenamefont {Angelani},\
  and\ \citenamefont {Iebba}}]{di2011swimming}%
  \BibitemOpen
  \bibfield  {author} {\bibinfo {author} {\bibfnamefont {R.}~\bibnamefont
  {Di~Leonardo}}, \bibinfo {author} {\bibfnamefont {D.}~\bibnamefont
  {Dell’Arciprete}}, \bibinfo {author} {\bibfnamefont {L.}~\bibnamefont
  {Angelani}}, \ and\ \bibinfo {author} {\bibfnamefont {V.}~\bibnamefont
  {Iebba}},\ }\bibfield  {title} {\enquote {\bibinfo {title} {Swimming with an
  image},}\ }\href@noop {} {\bibfield  {journal} {\bibinfo  {journal} {Physical
  review letters}\ }\textbf {\bibinfo {volume} {106}},\ \bibinfo {pages}
  {038101} (\bibinfo {year} {2011})}\BibitemShut {NoStop}%
\bibitem [{\citenamefont {Gaffney}\ \emph {et~al.}(2011)\citenamefont
  {Gaffney}, \citenamefont {Gad{\^e}lha}, \citenamefont {Smith}, \citenamefont
  {Blake},\ and\ \citenamefont {Kirkman-Brown}}]{gaffney2011mammalian}%
  \BibitemOpen
  \bibfield  {author} {\bibinfo {author} {\bibfnamefont {E.~A.}\ \bibnamefont
  {Gaffney}}, \bibinfo {author} {\bibfnamefont {H.}~\bibnamefont
  {Gad{\^e}lha}}, \bibinfo {author} {\bibfnamefont {D.~J.}\ \bibnamefont
  {Smith}}, \bibinfo {author} {\bibfnamefont {J.~R.}\ \bibnamefont {Blake}}, \
  and\ \bibinfo {author} {\bibfnamefont {J.~C.}\ \bibnamefont
  {Kirkman-Brown}},\ }\bibfield  {title} {\enquote {\bibinfo {title} {Mammalian
  sperm motility: observation and theory},}\ }\href@noop {} {\bibfield
  {journal} {\bibinfo  {journal} {Annual Review of Fluid Mechanics}\ }
  (\bibinfo {year} {2011})}\BibitemShut {NoStop}%
\bibitem [{\citenamefont {Feng}\ \emph {et~al.}(2022)\citenamefont {Feng},
  \citenamefont {Molina}, \citenamefont {Turner},\ and\ \citenamefont
  {Yamamoto}}]{feng2022dynamics}%
  \BibitemOpen
  \bibfield  {author} {\bibinfo {author} {\bibfnamefont {C.}~\bibnamefont
  {Feng}}, \bibinfo {author} {\bibfnamefont {J.~J.}\ \bibnamefont {Molina}},
  \bibinfo {author} {\bibfnamefont {M.~S.}\ \bibnamefont {Turner}}, \ and\
  \bibinfo {author} {\bibfnamefont {R.}~\bibnamefont {Yamamoto}},\ }\bibfield
  {title} {\enquote {\bibinfo {title} {Dynamics of microswimmers near a soft
  penetrable interface},}\ }\href@noop {} {\bibfield  {journal} {\bibinfo
  {journal} {arXiv preprint arXiv:2205.10919}\ } (\bibinfo {year}
  {2022})}\BibitemShut {NoStop}%
\bibitem [{\citenamefont {Lighthill}(1952)}]{Lighthill}%
  \BibitemOpen
  \bibfield  {author} {\bibinfo {author} {\bibfnamefont {M.~J.}\ \bibnamefont
  {Lighthill}},\ }\bibfield  {title} {\enquote {\bibinfo {title} {On the
  squirming motion of nearly spherical deformable bodies through liquids at
  very small reynolds numbers},}\ }\href {\doibase
  https://doi.org/10.1002/cpa.3160050201} {\bibfield  {journal} {\bibinfo
  {journal} {Communications on Pure and Applied Mathematics}\ }\textbf
  {\bibinfo {volume} {5}},\ \bibinfo {pages} {109--118} (\bibinfo {year}
  {1952})},\ \Eprint
  {http://arxiv.org/abs/https://onlinelibrary.wiley.com/doi/pdf/10.1002/cpa.3160050201}
  {https://onlinelibrary.wiley.com/doi/pdf/10.1002/cpa.3160050201} \BibitemShut
  {NoStop}%
\bibitem [{\citenamefont {Downton}\ and\ \citenamefont
  {Stark}(2009)}]{downton2009simulation}%
  \BibitemOpen
  \bibfield  {author} {\bibinfo {author} {\bibfnamefont {M.~T.}\ \bibnamefont
  {Downton}}\ and\ \bibinfo {author} {\bibfnamefont {H.}~\bibnamefont
  {Stark}},\ }\bibfield  {title} {\enquote {\bibinfo {title} {Simulation of a
  model microswimmer},}\ }\href@noop {} {\bibfield  {journal} {\bibinfo
  {journal} {Journal of Physics: Condensed Matter}\ }\textbf {\bibinfo {volume}
  {21}},\ \bibinfo {pages} {204101} (\bibinfo {year} {2009})}\BibitemShut
  {NoStop}%
\bibitem [{\citenamefont {Pak}\ and\ \citenamefont
  {Lauga}(2014)}]{pak2014generalized}%
  \BibitemOpen
  \bibfield  {author} {\bibinfo {author} {\bibfnamefont {O.~S.}\ \bibnamefont
  {Pak}}\ and\ \bibinfo {author} {\bibfnamefont {E.}~\bibnamefont {Lauga}},\
  }\bibfield  {title} {\enquote {\bibinfo {title} {Generalized squirming motion
  of a sphere},}\ }\href@noop {} {\bibfield  {journal} {\bibinfo  {journal}
  {Journal of Engineering Mathematics}\ }\textbf {\bibinfo {volume} {88}},\
  \bibinfo {pages} {1--28} (\bibinfo {year} {2014})}\BibitemShut {NoStop}%
\bibitem [{\citenamefont {Ishikawa}, \citenamefont {Simmonds},\ and\
  \citenamefont {Pedley}(2006)}]{ishikawa2006hydrodynamic}%
  \BibitemOpen
  \bibfield  {author} {\bibinfo {author} {\bibfnamefont {T.}~\bibnamefont
  {Ishikawa}}, \bibinfo {author} {\bibfnamefont {M.}~\bibnamefont {Simmonds}},
  \ and\ \bibinfo {author} {\bibfnamefont {T.~J.}\ \bibnamefont {Pedley}},\
  }\bibfield  {title} {\enquote {\bibinfo {title} {Hydrodynamic interaction of
  two swimming model micro-organisms},}\ }\href@noop {} {\bibfield  {journal}
  {\bibinfo  {journal} {Journal of Fluid Mechanics}\ }\textbf {\bibinfo
  {volume} {568}},\ \bibinfo {pages} {119--160} (\bibinfo {year}
  {2006})}\BibitemShut {NoStop}%
\bibitem [{\citenamefont {Arai}\ \emph {et~al.}(2020)\citenamefont {Arai},
  \citenamefont {Watanabe}, \citenamefont {Miyahara}, \citenamefont {Yamamoto},
  \citenamefont {Hampel},\ and\ \citenamefont {Lecrivain}}]{arai2020direct}%
  \BibitemOpen
  \bibfield  {author} {\bibinfo {author} {\bibfnamefont {N.}~\bibnamefont
  {Arai}}, \bibinfo {author} {\bibfnamefont {S.}~\bibnamefont {Watanabe}},
  \bibinfo {author} {\bibfnamefont {M.~T.}\ \bibnamefont {Miyahara}}, \bibinfo
  {author} {\bibfnamefont {R.}~\bibnamefont {Yamamoto}}, \bibinfo {author}
  {\bibfnamefont {U.}~\bibnamefont {Hampel}}, \ and\ \bibinfo {author}
  {\bibfnamefont {G.}~\bibnamefont {Lecrivain}},\ }\bibfield  {title} {\enquote
  {\bibinfo {title} {Direct observation of the attachment behavior of
  hydrophobic colloidal particles onto a bubble surface},}\ }\href@noop {}
  {\bibfield  {journal} {\bibinfo  {journal} {Soft matter}\ }\textbf {\bibinfo
  {volume} {16}},\ \bibinfo {pages} {695--702} (\bibinfo {year}
  {2020})}\BibitemShut {NoStop}%
\bibitem [{\citenamefont {Lecrivain}\ \emph {et~al.}(2020)\citenamefont
  {Lecrivain}, \citenamefont {Grein}, \citenamefont {Yamamoto}, \citenamefont
  {Hampel},\ and\ \citenamefont {Taniguchi}}]{lecrivain2020eulerian}%
  \BibitemOpen
  \bibfield  {author} {\bibinfo {author} {\bibfnamefont {G.}~\bibnamefont
  {Lecrivain}}, \bibinfo {author} {\bibfnamefont {T.~B.~P.}\ \bibnamefont
  {Grein}}, \bibinfo {author} {\bibfnamefont {R.}~\bibnamefont {Yamamoto}},
  \bibinfo {author} {\bibfnamefont {U.}~\bibnamefont {Hampel}}, \ and\ \bibinfo
  {author} {\bibfnamefont {T.}~\bibnamefont {Taniguchi}},\ }\bibfield  {title}
  {\enquote {\bibinfo {title} {Eulerian/lagrangian formulation for the
  elasto-capillary deformation of a flexible fibre},}\ }\href@noop {}
  {\bibfield  {journal} {\bibinfo  {journal} {Journal of Computational
  Physics}\ }\textbf {\bibinfo {volume} {409}},\ \bibinfo {pages} {109324}
  (\bibinfo {year} {2020})}\BibitemShut {NoStop}%
\bibitem [{\citenamefont {Yamamoto}, \citenamefont {Molina},\ and\
  \citenamefont {Nakayama}(2021)}]{yamamoto2021smoothed}%
  \BibitemOpen
  \bibfield  {author} {\bibinfo {author} {\bibfnamefont {R.}~\bibnamefont
  {Yamamoto}}, \bibinfo {author} {\bibfnamefont {J.~J.}\ \bibnamefont
  {Molina}}, \ and\ \bibinfo {author} {\bibfnamefont {Y.}~\bibnamefont
  {Nakayama}},\ }\bibfield  {title} {\enquote {\bibinfo {title} {Smoothed
  profile method for direct numerical simulations of hydrodynamically
  interacting particles},}\ }\href@noop {} {\bibfield  {journal} {\bibinfo
  {journal} {Soft Matter}\ }\textbf {\bibinfo {volume} {17}},\ \bibinfo {pages}
  {4226--4253} (\bibinfo {year} {2021})}\BibitemShut {NoStop}%
\bibitem [{\citenamefont {Yamamoto}, \citenamefont {Nakayama},\ and\
  \citenamefont {Kim}(2004)}]{yamamoto2004smooth}%
  \BibitemOpen
  \bibfield  {author} {\bibinfo {author} {\bibfnamefont {R.}~\bibnamefont
  {Yamamoto}}, \bibinfo {author} {\bibfnamefont {Y.}~\bibnamefont {Nakayama}},
  \ and\ \bibinfo {author} {\bibfnamefont {K.}~\bibnamefont {Kim}},\ }\bibfield
   {title} {\enquote {\bibinfo {title} {A smooth interface method for
  simulating liquid crystal colloid dispersions},}\ }\href@noop {} {\bibfield
  {journal} {\bibinfo  {journal} {Journal of Physics: Condensed Matter}\
  }\textbf {\bibinfo {volume} {16}},\ \bibinfo {pages} {S1945} (\bibinfo {year}
  {2004})}\BibitemShut {NoStop}%
\bibitem [{\citenamefont {Nakayama}\ and\ \citenamefont
  {Yamamoto}(2005)}]{PhysRevE.71.036707}%
  \BibitemOpen
  \bibfield  {author} {\bibinfo {author} {\bibfnamefont {Y.}~\bibnamefont
  {Nakayama}}\ and\ \bibinfo {author} {\bibfnamefont {R.}~\bibnamefont
  {Yamamoto}},\ }\bibfield  {title} {\enquote {\bibinfo {title} {Simulation
  method to resolve hydrodynamic interactions in colloidal dispersions},}\
  }\href {\doibase 10.1103/PhysRevE.71.036707} {\bibfield  {journal} {\bibinfo
  {journal} {Phys. Rev. E}\ }\textbf {\bibinfo {volume} {71}},\ \bibinfo
  {pages} {036707} (\bibinfo {year} {2005})}\BibitemShut {NoStop}%
\bibitem [{\citenamefont {Molina}\ and\ \citenamefont
  {Yamamoto}(2013)}]{molina2013direct}%
  \BibitemOpen
  \bibfield  {author} {\bibinfo {author} {\bibfnamefont {J.~J.}\ \bibnamefont
  {Molina}}\ and\ \bibinfo {author} {\bibfnamefont {R.}~\bibnamefont
  {Yamamoto}},\ }\bibfield  {title} {\enquote {\bibinfo {title} {Direct
  numerical simulations of rigid body dispersions. i. mobility/friction tensors
  of assemblies of spheres},}\ }\href@noop {} {\bibfield  {journal} {\bibinfo
  {journal} {The Journal of chemical physics}\ }\textbf {\bibinfo {volume}
  {139}},\ \bibinfo {pages} {234105} (\bibinfo {year} {2013})}\BibitemShut
  {NoStop}%
\bibitem [{\citenamefont {Datt}\ and\ \citenamefont
  {Elfring}(2019)}]{datt2019active}%
  \BibitemOpen
  \bibfield  {author} {\bibinfo {author} {\bibfnamefont {C.}~\bibnamefont
  {Datt}}\ and\ \bibinfo {author} {\bibfnamefont {G.~J.}\ \bibnamefont
  {Elfring}},\ }\bibfield  {title} {\enquote {\bibinfo {title} {Active
  particles in viscosity gradients},}\ }\href@noop {} {\bibfield  {journal}
  {\bibinfo  {journal} {Physical Review Letters}\ }\textbf {\bibinfo {volume}
  {123}},\ \bibinfo {pages} {158006} (\bibinfo {year} {2019})}\BibitemShut
  {NoStop}%
\bibitem [{\citenamefont {Liebchen}\ \emph {et~al.}(2018)\citenamefont
  {Liebchen}, \citenamefont {Monderkamp}, \citenamefont {Ten~Hagen},\ and\
  \citenamefont {L{\"o}wen}}]{liebchen2018viscotaxis}%
  \BibitemOpen
  \bibfield  {author} {\bibinfo {author} {\bibfnamefont {B.}~\bibnamefont
  {Liebchen}}, \bibinfo {author} {\bibfnamefont {P.}~\bibnamefont
  {Monderkamp}}, \bibinfo {author} {\bibfnamefont {B.}~\bibnamefont
  {Ten~Hagen}}, \ and\ \bibinfo {author} {\bibfnamefont {H.}~\bibnamefont
  {L{\"o}wen}},\ }\bibfield  {title} {\enquote {\bibinfo {title} {Viscotaxis:
  Microswimmer navigation in viscosity gradients},}\ }\href@noop {} {\bibfield
  {journal} {\bibinfo  {journal} {Physical review letters}\ }\textbf {\bibinfo
  {volume} {120}},\ \bibinfo {pages} {208002} (\bibinfo {year}
  {2018})}\BibitemShut {NoStop}%
\bibitem [{\citenamefont {Dandekar}\ and\ \citenamefont
  {Ardekani}(2020)}]{dandekar2020swimming}%
  \BibitemOpen
  \bibfield  {author} {\bibinfo {author} {\bibfnamefont {R.}~\bibnamefont
  {Dandekar}}\ and\ \bibinfo {author} {\bibfnamefont {A.~M.}\ \bibnamefont
  {Ardekani}},\ }\bibfield  {title} {\enquote {\bibinfo {title} {Swimming sheet
  in a viscosity-stratified fluid},}\ }\href@noop {} {\bibfield  {journal}
  {\bibinfo  {journal} {Journal of Fluid Mechanics}\ }\textbf {\bibinfo
  {volume} {895}} (\bibinfo {year} {2020})}\BibitemShut {NoStop}%
\bibitem [{\citenamefont {Eastham}\ and\ \citenamefont
  {Shoele}(2020)}]{eastham2020axisymmetric}%
  \BibitemOpen
  \bibfield  {author} {\bibinfo {author} {\bibfnamefont {P.~S.}\ \bibnamefont
  {Eastham}}\ and\ \bibinfo {author} {\bibfnamefont {K.}~\bibnamefont
  {Shoele}},\ }\bibfield  {title} {\enquote {\bibinfo {title} {Axisymmetric
  squirmers in stokes fluid with nonuniform viscosity},}\ }\href@noop {}
  {\bibfield  {journal} {\bibinfo  {journal} {Physical Review Fluids}\ }\textbf
  {\bibinfo {volume} {5}},\ \bibinfo {pages} {063102} (\bibinfo {year}
  {2020})}\BibitemShut {NoStop}%
\bibitem [{\citenamefont {Gidituri}\ \emph {et~al.}(2022)\citenamefont
  {Gidituri}, \citenamefont {Shen}, \citenamefont {W{\"u}rger},\ and\
  \citenamefont {Lintuvuori}}]{gidituri2022reorientation}%
  \BibitemOpen
  \bibfield  {author} {\bibinfo {author} {\bibfnamefont {H.}~\bibnamefont
  {Gidituri}}, \bibinfo {author} {\bibfnamefont {Z.}~\bibnamefont {Shen}},
  \bibinfo {author} {\bibfnamefont {A.}~\bibnamefont {W{\"u}rger}}, \ and\
  \bibinfo {author} {\bibfnamefont {J.~S.}\ \bibnamefont {Lintuvuori}},\
  }\bibfield  {title} {\enquote {\bibinfo {title} {Reorientation dynamics of
  microswimmers at fluid-fluid interfaces},}\ }\href@noop {} {\bibfield
  {journal} {\bibinfo  {journal} {Physical Review Fluids}\ }\textbf {\bibinfo
  {volume} {7}},\ \bibinfo {pages} {L042001} (\bibinfo {year}
  {2022})}\BibitemShut {NoStop}%
\bibitem [{\citenamefont {Daniels}, \citenamefont {Longland},\ and\
  \citenamefont {Gilbart}(1980)}]{daniels1980aspects}%
  \BibitemOpen
  \bibfield  {author} {\bibinfo {author} {\bibfnamefont {M.~J.}\ \bibnamefont
  {Daniels}}, \bibinfo {author} {\bibfnamefont {J.~M.}\ \bibnamefont
  {Longland}}, \ and\ \bibinfo {author} {\bibfnamefont {J.}~\bibnamefont
  {Gilbart}},\ }\bibfield  {title} {\enquote {\bibinfo {title} {Aspects of
  motility and chemotaxis in spiroplasmas},}\ }\href@noop {} {\bibfield
  {journal} {\bibinfo  {journal} {Microbiology}\ }\textbf {\bibinfo {volume}
  {118}},\ \bibinfo {pages} {429--436} (\bibinfo {year} {1980})}\BibitemShut
  {NoStop}%
\bibitem [{\citenamefont {Takabe}\ \emph {et~al.}(2017)\citenamefont {Takabe},
  \citenamefont {Tahara}, \citenamefont {Islam}, \citenamefont {Affroze},
  \citenamefont {Kudo},\ and\ \citenamefont {Nakamura}}]{takabe2017viscosity}%
  \BibitemOpen
  \bibfield  {author} {\bibinfo {author} {\bibfnamefont {K.}~\bibnamefont
  {Takabe}}, \bibinfo {author} {\bibfnamefont {H.}~\bibnamefont {Tahara}},
  \bibinfo {author} {\bibfnamefont {M.~S.}\ \bibnamefont {Islam}}, \bibinfo
  {author} {\bibfnamefont {S.}~\bibnamefont {Affroze}}, \bibinfo {author}
  {\bibfnamefont {S.}~\bibnamefont {Kudo}}, \ and\ \bibinfo {author}
  {\bibfnamefont {S.}~\bibnamefont {Nakamura}},\ }\bibfield  {title} {\enquote
  {\bibinfo {title} {Viscosity-dependent variations in the cell shape and
  swimming manner of leptospira},}\ }\href@noop {} {\bibfield  {journal}
  {\bibinfo  {journal} {Microbiology}\ }\textbf {\bibinfo {volume} {163}},\
  \bibinfo {pages} {153--160} (\bibinfo {year} {2017})}\BibitemShut {NoStop}%
\bibitem [{\citenamefont {Coppola}\ and\ \citenamefont
  {Kantsler}(2021)}]{coppola2021green}%
  \BibitemOpen
  \bibfield  {author} {\bibinfo {author} {\bibfnamefont {S.}~\bibnamefont
  {Coppola}}\ and\ \bibinfo {author} {\bibfnamefont {V.}~\bibnamefont
  {Kantsler}},\ }\bibfield  {title} {\enquote {\bibinfo {title} {Green algae
  scatter off sharp viscosity gradients},}\ }\href@noop {} {\bibfield
  {journal} {\bibinfo  {journal} {Scientific reports}\ }\textbf {\bibinfo
  {volume} {11}},\ \bibinfo {pages} {1--7} (\bibinfo {year}
  {2021})}\BibitemShut {NoStop}%
\bibitem [{\citenamefont {Stehnach}\ \emph {et~al.}(2021)\citenamefont
  {Stehnach}, \citenamefont {Waisbord}, \citenamefont {Walkama},\ and\
  \citenamefont {Guasto}}]{stehnach2021viscophobic}%
  \BibitemOpen
  \bibfield  {author} {\bibinfo {author} {\bibfnamefont {M.~R.}\ \bibnamefont
  {Stehnach}}, \bibinfo {author} {\bibfnamefont {N.}~\bibnamefont {Waisbord}},
  \bibinfo {author} {\bibfnamefont {D.~M.}\ \bibnamefont {Walkama}}, \ and\
  \bibinfo {author} {\bibfnamefont {J.~S.}\ \bibnamefont {Guasto}},\ }\bibfield
   {title} {\enquote {\bibinfo {title} {Viscophobic turning dictates microalgae
  transport in viscosity gradients},}\ }\href@noop {} {\bibfield  {journal}
  {\bibinfo  {journal} {Nature Physics}\ }\textbf {\bibinfo {volume} {17}},\
  \bibinfo {pages} {926--930} (\bibinfo {year} {2021})}\BibitemShut {NoStop}%
\bibitem [{\citenamefont {L{\'o}pez}\ \emph {et~al.}(2021)\citenamefont
  {L{\'o}pez}, \citenamefont {Gonzalez-Gutierrez}, \citenamefont
  {Solorio-Ordaz}, \citenamefont {Lauga},\ and\ \citenamefont
  {Zenit}}]{lopez2021dynamics}%
  \BibitemOpen
  \bibfield  {author} {\bibinfo {author} {\bibfnamefont {C.~E.}\ \bibnamefont
  {L{\'o}pez}}, \bibinfo {author} {\bibfnamefont {J.}~\bibnamefont
  {Gonzalez-Gutierrez}}, \bibinfo {author} {\bibfnamefont {F.}~\bibnamefont
  {Solorio-Ordaz}}, \bibinfo {author} {\bibfnamefont {E.}~\bibnamefont
  {Lauga}}, \ and\ \bibinfo {author} {\bibfnamefont {R.}~\bibnamefont
  {Zenit}},\ }\bibfield  {title} {\enquote {\bibinfo {title} {Dynamics of a
  helical swimmer crossing viscosity gradients},}\ }\href@noop {} {\bibfield
  {journal} {\bibinfo  {journal} {Physical Review Fluids}\ }\textbf {\bibinfo
  {volume} {6}},\ \bibinfo {pages} {083102} (\bibinfo {year}
  {2021})}\BibitemShut {NoStop}%
\end{thebibliography}%

\end{document}